\documentclass[12pt, a4paper]{article}
\usepackage{mathrsfs}
\usepackage{bbm}
\usepackage{amsmath,amssymb}
\usepackage{color}

\usepackage[titletoc]{appendix}

\allowdisplaybreaks \numberwithin{equation}{section}
\numberwithin{equation}{section} 

 \newtheorem{thm}{Theorem}[section]
 \newtheorem{cor}[thm]{Corollary}
 \newtheorem{lem}[thm]{Lemma}
 \newtheorem{prop}[thm]{Proposition}
 \newtheorem{exam}[thm]{Example}
 \newtheorem{defn}[thm]{Definition}
 \newtheorem{rem}[thm]{Remark}
\newenvironment{prf}{\noindent {\it Proof} \ }{\hfill $\Box$}
\newenvironment{prfn}[1]{\noindent {\it Proof of #1} \ }{\hfill $\Box$}

\newcommand{\eqa}{\begin{eqnarray}}
\newcommand{\eeqa}{\end{eqnarray}}
\newcommand{\beq}{\begin{equation}}
\newcommand{\eeq}{\end{equation}}

\newcommand\p{\partial}
\newcommand{\bm}[1]{\mbox{\boldmath{$#1$}}}

\newcommand{\nn}{\nonumber}
\newcommand{\ve}{\varepsilon}
\newcommand\ep{\epsilon}
\newcommand{\lm}{\lambda}
\newcommand{\al}{\alpha}
\newcommand{\gm}{\gamma}
\newcommand{\sg}{\sigma}
\newcommand{\om}{\omega} 
\newcommand{\Gm}{\Gamma}

\newcommand{\dt}{\delta}
\newcommand{\ta}{\theta}

\newcommand{\res}{\mathrm{res}}

\newcommand\bmi{\mathbbm{i}}

\newcommand\ra{\rangle}
\newcommand\la{\langle}

\newcommand{\cM}{\mathcal{M}}

\newcommand{\mathZ}{\mathbb{Z}}

\newcommand{\mathH}{\mathcal{H}}

\newcommand{\mathL}{\mathcal{L}}
\newcommand{\mathM}{\mathcal{M}}

\newcommand{\mathT}{{T}}
\newcommand{\cF}{\mathcal{F}}
\newcommand{\mathS}{{S}}

\begin{document}
\title{A Class of Infinite-dimensional Frobenius Manifolds and Their Submanifolds}
\author{
Chao-Zhong Wu
\qquad Dingdian Xu
\\
\\
{\small $^*$SISSA, Via Bonomea 265, 34136 Trieste, Italy}
\\
{\small $^\dag$Department of Mathematical Sciences, Tsinghua
University, Beijing 100084, P.R. China} }

\date{}
\maketitle

\begin{abstract}
 We construct a class of
 infinite-dimensional Frobenius manifolds on the space of
 pairs of certain even functions meromorphic inside or outside
 the unit circle. Via a bi-Hamiltonian recursion relation, the
 principal hierarchies associated to such Frobenius manifolds are
 found to be certain extensions of the dispersionless two-component BKP
 hierarchy. Moreover, we show that these
 manifolds contain finite-dimensional Frobenius
 submanifolds as defined on the orbit space of Coxeter
 groups of types B and D.
\end{abstract}

\section{Introduction}

The notion of Frobenius manifold was introduced by Dubrovin
\cite{Du} to give a geometric
  description of the WDVV equation in the context of 2D topological field
  theory  \cite{DVV, Witten}. This concept was discovered to be essential
  in characterizing a wide class of integrable systems that are related
  to branches of mathematical physics such as topological field theory,
  Gromov-Witten invariants and singularity theory, see \cite{Du, DLZ, DZ,
 Givental, GM} and references therein.

   Recall that a \emph{Frobenius algebra} $(A, e, <\ ,\ >)$  is a
   commutative associative algebra  $A$ with a unity $e$ and
   a non-degenerate symmetric bilinear form (inner product) $<\ ,\ >$ that is invariant with
   respect to the multiplication, i.e.,
   \begin{align}
     < x\cdot y,z>=< x,y\cdot z>, \quad  x,y,z\in A.
   \end{align}
 \begin{defn}[\cite{Du}]
   A manifold $M$ is called a Frobenius manifold if
   on each tangent
     space $T_t M$ a Frobenius algebra $(T_t M, e, <\ ,\ >)$  is defined
     depending smoothly on $t\in M$, and the following axioms are fulfilled:
  \begin{description}
     \item[(F1)] The inner product $<\ ,\ >$ is a flat metric on
     $M$. Denote the Levi-Civita connection for this metric by $\nabla$, then the
     unity vector field $e$ satisfies $\nabla e=0$.
     \item[(F2)] Let $c$ be the 3-tensor $c(x,y,z):=<x\cdot y,z>$, then the
     4-tensor $(\nabla_{w}c)(x,y,z)$
     is symmetric in the vector fields $x,y,z,w$.
     \item[(F3)] A so-called Euler vector field $E\in Vect(M)$ is fixed on M
     such that $\nabla\nabla E=0$, and it satisfies
     \begin{align}
        [E,x\cdot y]-[E,x]\cdot y-x\cdot [E,y]&=x\cdot y,\nn\\
        Lie_E <x,y>-<[E,x],y>-<x,[E,y]>&=(2-d)<x,y>.\nn
    \end{align}
    Here $d$ is a constant named as the charge of $M$.
  \end{description}
   \end{defn}

   On a Frobenius manifold $M$ there exist \emph{flat coordinates}
  $t^1,\dots,t^n$ such that the unity vector field $e=\p/\p t^1$ and
   \begin{align}
     (\eta_{\al\beta})_{n\times n}=\left(<\frac{\p}{\p t^\al},\frac{\p}{\p
     t^{\beta}}>\right)_{n\times n}
   \end{align}
   is a constant non-degenerate matrix. Denote the inverse of this
   matrix by $(\eta^{\al\beta})_{n\times n}$, then the product of the Frobenius algebra
   $T_t M$ reads
   \begin{align}
     \frac{\p}{\p t^{\al}}\cdot \frac{\p}{\p t^{\beta}}=c^{\gm}_{\al\beta}\frac{\p}{\p
     t^{\gm}}, \quad c^{\gm}_{\al\beta}=\eta^{\gm\ep}c_{\ep\al\beta}
   \end{align}
with $c_{\al\beta\gm}=c(\p/\p t^\al,\p/\p t^\beta,\p/\p t^\gm)$.
Here the Einstein convention of summation over repeated indices is
   used.

Note that the structure constants satisfy
\begin{equation}\label{WDVV1}
c_{1\al}^{\beta}=\dt_\al^\beta, \quad c_{\al\beta}^\ep
c_{\ep\gm}^{\sg}=c_{\al\gm}^\ep c_{\ep\beta}^{\sg},
\end{equation}
and there locally exists on $M$ a smooth function $F$, called the
\emph{potential} of the Frobenius manifold, such that
\begin{equation}\label{WDVV2}
c_{\al\beta\gm}=\frac{\p^3 F}{\p t^{\al}\p t^{\beta}\p t^{\gm}},
\quad Lie_E F=(3-d)F+\hbox{ quadratic}.
\end{equation}
The system \eqref{WDVV1}, \eqref{WDVV2} is called the WDVV equation
in topological field theory \cite{DVV, Witten}, and its solution $F$
is the free energy whose third derivatives $c_{\al\beta\gm}$ are the
$3$-point correlator functions.

Conversely, suppose it is given a potential $F$ satisfying
\eqref{WDVV1} and \eqref{WDVV2} (which also implies a constant
matrix $(\eta_{\al\beta})$, a unity and a Euler vector field), then
one can recover the structure of a Frobenius manifold.

A Frobenius manifold $M$ is said to be \emph{semisimple} if the
Frobenius algebras $T_t M$ are semisimple at generic points $t\in
M$. On a semisimple Frobenius manifold there exist so-called
\emph{canonical coordinates} that are given by the eigenvalues of
the operator of multiplication by the Euler vector field \cite{Du,
Du2}.

Due to the associativity property \eqref{WDVV1}, there is a
commutative associative algebra structure on every cotangent space
$T_t^*M$ of a Frobenius manifold $M$. More precisely,  the structure
constants for the basis $\{d t^{\al}\}$ are
$c^{\al\beta}_\gm=\eta^{\al\ep}c_{\ep\gm}^\beta$. Hence on $T_t^*M$
a symmetric bilinear form, named the \emph{intersection form}, is
defined by
   \begin{align}\label{inters}
     (dt^{\al},dt^{\beta})^*:=i_{E}(dt^{\al}\cdot dt^{\beta}).
   \end{align}
A celebrated result is that this intersection form and the bilinear
form $<dt^\al,dt^\beta>^*=\eta^{\al\beta}$ compose a flat pencil of
metrics on $T_t^*M$. That is, every linear combination (with
constant coefficients) of the two metrics is a flat metric, and the
Christoffel symbols $\Gamma^{\al\beta}_{\gamma}$ of their
Levi-Civita connections obey the same linear combination relation.

Every flat pencil of metrics defines a local bi-Hamiltonian
structure of hydrodynamic type \cite{DN}, which induces an
integrable system of bi-Hamiltonian equations. Based on this fact,
on the loop space $\left\{S^1\to M\right\}$ of the Frobenius
manifold $M$ there is an integrable hierarchy carrying a
bi-Hamiltonian structure of hydrodynamic type. Such a hierarchy is
called the \emph{principal hierarchy} \cite{Du, DZ}, in which the
evolutionary equations of the lowest level, i.e., the \emph{primary
flows}, read
\begin{equation}\label{prflow}
\frac{\p t^\gm}{\p T^{\al,0}}=c^{\gm}_{\al\beta}\frac{\p t^\beta}{\p
x}, \quad \al, \gm=1, 2, \dots, n
\end{equation}
with $x$ being the coordinate of $S^1$, and the flows of higher
level can be determined via a bi-Hamiltonian recursion relation with
Hamiltonian densities $\ta_{\al,p}(t)$ defined by
\begin{align}\label{recursion}
\ta_{\al,0}=\eta_{\al\beta}t^{\beta},\quad \ta_{\al,1}=\frac{\p
F}{\p t^{\al}}, \quad \frac{\p^2\ta_{\al,p}}{\p t^\lm\p
t^{\mu}}=c^{\ep}_{\lm\mu}\frac{\p\ta_{\al,p-1}}{\p t^{\ve}} \hbox{
for }  p>1.
\end{align}

Up to now various Frobenius manifolds of finite dimension have been
constructed. A typical example is the construction of semisimple
Frobenius manifolds on the orbit space of Coxeter groups \cite{Du3,
Du}. In Particular, for Frobenius manifolds on the orbit space of
Weyl groups, the principal hierarchies are the dispersionless limit
of Drinfeld-Sokolov hierarchies associated to untwisted affine
Kac-Moody algebras with the zeroth vertex $c_0$ of the Dynkin
diagram marked \cite{DS, Kac, Du, DLZ}, and they are closely related
to singularity theory and Gromov-Witten invariants \cite{DZ, GM}.
Another important class of Frobenius manifolds are defined on the
orbit space of the extended affine Weyl groups \cite{DZ2}, for part
of which the principal hierarchies are the dispersionless limit of
the extended bigraded Toda hierarchies \cite{Carlet, CDZ}.

The profound theory of finite-dimensional Frobenius manifold and
integrable hierarchies has started recently being generalized to the
case of infinite dimension. The first infinite-dimensional Frobenius
manifold was constructed by Carlet, Dubrovin and Mertens \cite{CDM}
in consideration of a bi-Hamiltonian structure of the dispersionless
2D Toda hierarchy. The other published infinite-dimensional
Frobenius manifold, proposed by Raimondo \cite{Raimondo}, underlies
the dispersionless KP hierarchy.

It is well known that the KP hierarchy and the 2D Toda hierarchy can
be reduced to the Gelfand-Dickey hierarchies and the extended
bigraded Toda hierarchies (except the flows defined by the logarithm
operator) respectively \cite{Dickey, CDZ, Carlet}. These reduced
hierarchies are deformations of the principal hierarchy for the
Frobenius manifolds on the orbit space of (extended) Coxeter groups
of type A. Such reductions of integrable hierarchies as well as the
bi-Hamiltonian structures are expected to be described by certain
finite-dimensional Frobenius submanifolds of the
infinite-dimensional ones. For instance, in \cite{CDM} a submanifold
of two dimensions was considered, whose principal hierarchy is the
dispersionless extended Toda hierarchy \cite{CDZ}, i.e., the
extended bigraded Toda hierarchy \cite{Carlet} with the simplest
parameter $(1, 1)$.

In this paper we will follow the approach of \cite{CDM} and
construct a series of infinite-dimensional Frobenius manifolds. For
each of them, the principal hierarchy will be derived, which carries
a bi-Hamiltonian structure of the dispersionless two-component BKP
hierarchy  (see \cite{DJKM, Takasaki, LWZ} or Section~3 below for
the definition). Furthermore, corresponding to the reductions of
these bi-Hamiltonian structures studied in \cite{Wu}, we show that
there are finite-dimensional Frobenius submanifolds coinciding with
those defined on the orbit space of Coxeter groups of types B and D
\cite{Bertola, Du3, Du, Zuo}. In fact, it is the properties of such
finite-dimensional Frobenius manifolds that provide much inspiration
for us to come to a relatively neat description of the
infinite-dimensional manifolds.

Let us state the main results. Given $0<\ep<1$, we introduce two
sets of holomorphic even functions on disks of
$\mathbb{CP}^1=\mathbb{C}\cup\{\infty\}$ as follows:
 \begin{align}
     \mathH^-=\left\{f(z)=\sum_{i\ge0}f_i\,z^{-2 i}\mid f \hbox{ holomorphic on
     } |z|>1-\ep \right\},\nn\\
 \mathH^+=\left\{\hat{f}(z)=\sum_{i\ge0}\hat{f}_i\,z^{2 i}\mid \hat{f} \hbox{ holomorphic on
     } |z|<1+\ep \right\}.\nn
    \end{align}
For two arbitrary positive integers $m$ and $n$ we consider the
following coset
   \begin{align}
    \tilde{\cM}_{m,n}=(z^{2m},0)+z^{2m-2}\mathH^-\times z^{-2 n}
     \mathH^+.
   \end{align}
An element of this coset reads $\bm{a}(z)=\big(a(z),\hat{a}(z)\big)$
where
   \begin{align} \label{aah}
      a(z)=z^{2m}+\sum_{i\leq m}v_i z^{2i-2}, \quad
   \hat{a}(z)=\sum_{j\geq -n}\hat{v}_j z^{2j}.
   \end{align}
With coordinates given by the coefficients $v_i$ and $\hat{v}_j$ of
such Laurent series, the coset $ \tilde{\cM}_{m,n}$ is viewed as an
infinite-dimensional manifold. This manifold has tangent and
cotangent spaces as follows:
\begin{align}\label{tang}
  \mathT_{\bm{a}}\tilde{\cM}_{m,n}=z^{2m-2}\mathH^-\times z^{-2 n}
     \mathH^+, \quad
  \mathT_{\bm{a}}^*\tilde{\cM}_{m,n}=z^{-2m+1}\mathH^+\times z^{2 n-1}\mathH^-.
\end{align}
The pairing between a vector $\bm{\al}=(\al(z),\hat{\al}(z))\in
\mathT_{\bm{a}}\tilde{\cM}_{m,n}$ and a covector
$\bm{\om}=(\om(z),\hat{\om}(z))\in
\mathT^*_{\bm{a}}\tilde{\cM}_{m,n}$ is defined to be
\begin{align}\label{dualpair}
  \la\bm{\al}, \bm{\om}\ra=\frac{1}{2\pi\bmi}\oint_{|z|=1}\big[\al(z)\om(z)+\hat{\al}(z)\hat{\om}(z)\big]dz.
\end{align}
For $(a(z),\hat{a}(z))\in\tilde{\cM}_{m,n}$  we let
\begin{align}\label{funcpair}
  \zeta(z)=a(z)-\hat{a}(z),\quad  l(z)=a(z)_++\hat{a}(z)_-,
\end{align}
where the subscripts ``$\pm$'' mean the projections of a Laurent
series to its nonnegative part and negative part respectively.

Let $\cM_{m,n}$ be a submanifold of $\tilde{\cM}_{m,n}$ that
consists of points $(a(z),\hat{a}(z))$ satisfying the following
conditions:
\begin{description}
\item[(M1)] The function $\hat{a}(z)$ has a pole of order $2 n$ at $0$, namely,
$\hat{v}_{-n}\neq
0$;
\item[(M2)] For $|z|=1$,
\begin{align}\label{constraint}
 a(z)\hat{a}'(z)-a'(z)\hat{a}(z)\neq 0,
  \quad  \zeta'(z)\neq 0,\quad  l'(z)\neq 0;
\end{align}
\item[(M3)]
The winding number of the function $\zeta(z)$ around 0 is $2$, such
that $w(z)=\zeta(z)^{1/2}$ maps $S^1$
 biholomorphicly to a simple smooth curve $\Gamma$ around $0$.
\end{description}
The manifold $\cM_{m,n}$ is the one on which a Frobenius structure
will be constructed, and our discussion is restricted to it below.

Introduce variables $\bm{t}=\{t^i\}_{i\in\mathZ}$,
$\bm{h}=\big\{h^j\big\}_{j=1}^{m}$ and
$\hat{\bm{h}}=\big\{\hat{h}^k\big\}_{k=1}^{n}$ as
  \begin{align}
     t^i:=&\frac{2}{2i-1}\frac{1}{2\pi\bmi}\oint_{|z|=1}\zeta(z)^{\frac{-2i+1}{2}}dz,
     \quad i\in\mathZ;  \label{flatc01}
     \\
     h^j:=&-\frac{2m}{2j-1}\res_{z=\infty}~l(z)^{\frac{2j-1}{2m}}dz,
     \quad j=1,2,\cdots,m; \label{flatc02}\\
     \hat{h}^k:=&\frac{2n}{2k-1}\res_{z=0}~l(z)^{\frac{2k-1}{2n}}dz,
     \quad k=1,2,\cdots,n. \label{flatc03}
  \end{align}
One sees below that these variables give another system of
coordinates on the manifold $\cM_{m,n}$.

 \begin{thm}\label{main}
     For any two positive integers $m$ and $n$, the infinite dimensional manifold $\cM_{m,n}$
     is a semisimple Frobenius manifold of charge $d_m=1-\frac{1}{m}$ such that
    \begin{itemize}
      \item[(i)] The unity vector field $\bm{e}=\p/\p{h^m}$;
\item[(ii)] The Euler vector field
  \begin{align}\label{euler}
    \mathcal{E}_{m,n}=\sum_{i\in\mathZ}\frac{m(1-2i)+1}{2m}t^i\frac{\p}{\p t^i}+\sum_{j=1}^m\frac{j}{m}h^j\frac{\p}{\p h^j}
     +\sum_{k=1}^n\Big(\frac{2k-1}{2n}+\frac{1}{2m}\Big)\hat{h}^k\frac{\p}{\p\hat{h}^k};
  \end{align}
\item[(iii)] The potential
     \begin{align}\label{poten}
     \mathcal{F}_{m,n}=&\left(\frac{1}{2\pi\bmi}\right)^2
     \oint\oint_{|z_1|<|z_2|}\left(\frac{1}{2}\zeta(z_1)\zeta(z_2)-\zeta(z_1)l(z_2)
     +l(z_1)\zeta(z_2)\right)\times \nn \\
     & \qquad\qquad\qquad \times\log\left(\frac{z_2-z_1}{z_2}\right)\,dz_1dz_2+F_{m,n},
  \end{align}
  where $F_{m,n}$ is a rational function of variables
  $\bm{h}\cup\hat{\bm{h}}$ determined by
  \begin{align}\label{poten02}
    \frac{\p^3 F_{m,n}}{\p u\, \p v\, \p w}
    =-\big(\res_{z=\infty}+\res_{z=0}\big)\frac{\p_{u} l(z)\cdot\p_{v}
    l(z)\cdot\p_{w} l(z)}{l'(z)}dz
  \end{align}
  for any $u,v, w\in\bm{h}\cup\hat{\bm{h}}$.
    \end{itemize}
   \end{thm}
This theorem will be proved in Section~2, where a flat metric (with
flat coordinates \eqref{flatc01}--\eqref{flatc03}), an intersection
form and the canonical coordinates of this semisimple Frobenius
manifold will also be given. We remark that this Frobenius structure
is distinct from the one in \cite{CDM} restricted to the space
consisting of even Laurent series.

In section~3 we will show that the Frobenius manifold $\cM_{m,n}$ is
associated with a bi-Hamiltonian structure of hydrodynamic type,
which is the one derived in \cite{Wu} for the dispersionless
two-component BKP hierarchy, see also \cite{WX}. Then via the
bi-Hamiltonian recursion relation the principal hierarchy for
$\cM_{m,n}$ will be obtained.
\begin{thm}\label{prinhier}
Suppose the coefficients $v_i$ and $\hat{v}_j$ in \eqref{aah} are
smooth functions of $x\in S^1$. Then the principal hierarchy
associated to the Frobenius manifold $\cM_{m,n}$ can be written in
the following Lax form:
\begin{align} \label{principal}
  \frac{\p  a(z)}{\p T^{u,p}}=\left\{-(A_{u,p}(z))_-,  a(z)\right\},\quad
  \frac{\p\hat{ a}(z)}{\p
  T^{u,p}}=\left\{(A_{u,p}(z))_+,\hat{ a}(z)\right\}, \quad p\ge0,
  \end{align}
  where $\{f,g\}=\p f/\p z\cdot\p g/\p x-\p g/\p z\cdot\p f/\p
  x$ and
\begin{equation}\label{Aup}
 A_{u,p}(z)=\left
 \{ \begin{aligned}
    &\frac{1}{2 i+1}\frac{1}{(2 p)!!}\zeta(z)^{\frac{2i+1}{2}}\varphi(z)^p,
      &&u=t^i ~( i\in\mathbb{Z}); \\
       \\
    &\frac{\Gamma\left(\frac{2m-2j+1}{2m}\right)}{2m\,\Gamma\left(p+1+\frac{2m-2j+1}{2m}\right)}
a(z)^{p+\frac{2m-2j+1}{2m}}, && u=h^j ~ (1\le j\le m); \\
    \\
    &\frac{\Gamma\left(\frac{2n-2k+1}{2n}\right)}{2n\,\Gamma\left(p+1+\frac{2n-2k+1}{2n}\right)}
     \hat{a}(z)^{p+\frac{2n-2k+1}{2n}}, && u=\hat{h}^k ~ (1\le k\le
     n)
          \end{aligned}
          \right.
 \end{equation}
 with $\varphi(z)=a(z)+\hat{a}(z)$.
\end{thm}
Observe that the flows $\p/\p T^{h^j,p}$ and $\p/\p T^{\hat{h}^k,p}$
compose the dispersionless two-component BKP hierarchy. For this
reason we call the system of equations \eqref{principal} the
\emph{principal two-component BKP hierarchy}. Recall that this
hierarchy depends on two parameters $m$ and $n$.

In section~4 we will consider an $(m+n)$-dimensional submanifold
$M_{m,n}\subset\cM_{m,n}$ spanned by the flat coordinates
$\bm{h}\cup\hat{\bm{h}}$. The following proposition is a corollary
of the results of Sections~2 and~3.
\begin{prop} \label{prp-mn}
The submanifold $M_{m,n}$ is a semisimple Frobenius submanifold of
$\cM_{m,n}$, i.e., $M_{m,n}$ carries a Frobenius structure projected
from that of $\cM_{m,n}$. The potential $F_{m,n}$ (given in
\eqref{poten}) of $M_{m,n}$ is a polynomial in ($h^1$, $\dots$,
$h^{m}$, $\hat{h}^1$, $\dots$, $\hat{h}^n$, $1/\hat{h}^1$). It is a
polynomial in the flat coordinates ($h^1$, $\dots$, $h^{m}$,
$\hat{h}^1$, $\dots$, $\hat{h}^n$) if and only if $n=1$.

The principal hierarchy associated to $M_{m,n}$ is the
$(2m,2n)$-reduction of the dispersionless two-component BKP
hierarchy \cite{DJKM2, LWZ}, that is, the restriction of
\eqref{principal} to $a(z)=\hat{a}(z)=l(z)$. Particularly when
$n=1$, the principal hierarchy is the dispersionless limit of the
Drinfeld-Sokolov hierarchy of type $(D^{(1)}_{m+1},c_0)$.
\end{prop}

One will see that $M_{m,n}$ coincides with the Frobenius manifold
defined on the orbit space of the Coxeter group $B_{m+n}$ by Bertola
\cite{Bertola} with superpotential $l(z)$ given in \eqref{funcpair},
and constructed also by Zuo \cite{Zuo} in a different way. However,
to our best knowledge, the principal hierarchy for the Frobenius
manifold $M_{m,n}$ with general $(m,n)$ has not been considered
before.

The last section is devoted to the conclusion and outlook.

\section{Construction of Frobenius manifolds}

Let us begin to prove Theorem~\ref{main}, i.e., to equip a
semisimple Frobenius structure to the infinite-dimensional manifold
\[
\mathM_{m,n}\subset (z^{2m},0)+z^{2m-2}\mathH^-\times z^{-2 n}
     \mathH^+
\]
constrained by conditions (M1) -- (M3) above with any fixed positive
integers $m$ and $n$. To this end, we need to introduce a flat
metric that is invariant with respect to a multiplication on each
tangent space, and to write down the potential and the Euler vector
field. Besides that, the canonical coordinates and the intersection
form on this Frobenius manifold will be derived.

\subsection{A flat metric}

Recall that a system of coordinates of $\mathM_{m,n}$ is given by
the coefficients of the series
\[
\bm{a}=(a(z),\hat{a}(z))= \left(z^{2m}+\sum_{i\leq m}v_i z^{2i-2},
\sum_{j\geq -n}\hat{v}_j z^{2j}\right).
\]
We identify the tangent space $\mathT_{\bm{a}}\mathM_{m,n}$ with
$z^{2m-2}\mathH^-\times z^{-2n}\mathH^+$ by
\begin{align}
  \p=\ (\p a(z),\p \hat{a}(z))
\end{align}
for any tangent vector $\p$. For example, when $i\leq m$ and $j\geq
-n$ we have
\begin{equation}\label{dvvh}
\frac{\p}{\p v_i}=(z^{2i-2},0),\quad \frac{\p}{\p
\hat{v}_j}=(0,z^{2j}).
\end{equation}
According to the pairing \eqref{dualpair},
$\mathT^*_{\bm{a}}\mathM_{m,n}$ is identified with
$z^{-2m+1}\mathH^+\times z^{2n-1}\mathH^-$, and its dual basis with
respect to \eqref{dvvh} reads
\begin{align}\label{nacc}
  dv_i=(z^{-2i+1},0),\quad d\hat{v}_j=(0,z^{-2j-1})
\end{align}
with  $i\leq m$ and $j\geq -n$.

Introduce two generating functions for covectors
\begin{align}
  &da(p):=\sum_{i\leq m}d v_i p^{2i-2}=\Big(\frac{p^{2m}}{z^{2m-1}(p^2-z^2)},0\Big), \quad |z|<|p|,  \label{rpgener01} \\
  &d\hat{a}(p):=\sum_{j\geq -n}d\hat{v}_j p^{2j}=\Big(0,\frac{z^{2n+1}}{p^{2n}(z^2-p^2)}\Big), \quad |z|>|p|. \label{rpgener02}
\end{align}
The following lemma follows from the Cauchy integral formula.
\begin{lem}
The generating functions \eqref{rpgener01} and \eqref{rpgener02} have the following
properties:
\begin{itemize}
\item[(i)]
 For any vector $\bm{\xi}=(\xi(z),\hat{\xi}(z))\in\mathT_{\bm{a}}\cM_{m,n}$,
\begin{align}
  \la da(p),\bm{\xi}\ra=\xi(p),\quad \la d\hat{a}(p),\bm{\xi}\ra=\hat{\xi}(p);
\end{align}
\item[(ii)] For any covector
$\bm{\om}=(\om(z),\hat{\om}(z))\in\mathT^*_{\bm{a}}\mathM_{m,n}$,
\begin{align}
  &\bm{\om}=\frac{1}{2\pi\bmi}\oint_{|z|=1}\big[\om(p)da(p)+\hat{\om}(p)d\hat{a}(p)\big]d p.\label{gener}
\end{align}
\end{itemize}

\end{lem}

Let us introduce a symmetric bilinear form on the cotangent space
$\mathT^*_{\bm{a}}\mathM_{m,n}$ as
\begin{align}\label{cometric}
  &<d\al(p),d\beta(q)>^*=\frac{q\beta'(q)}{q^2-p^2}+\frac{p\al'(p)}{p^2-q^2},
\end{align}
where $\al'(p)={\p\al(p)}/{\p p}$ and $\al,\beta\in\{a,\hat{a}\}$.
Note the difference between the above formula and equation~(1.16) in
\cite{CDM}.

Since the pairing \eqref{dualpair} is nondegenerate, the following
linear map is well defined
\begin{align}\label{linmap}
\eta:\ \mathT_{\bm{a}}^*\mathM_{m,n}\rightarrow
\mathT_{\bm{a}}\mathM_{m,n}, \quad \la
\bm{\om}_1,\eta(\bm{\om}_2)\ra=<\bm{\om}_1,\bm{\om}_2>^*
\end{align}
for any $\bm{\om}_1,\bm{\om}_2\in \mathT_{\bm{a}}^* \mathM_{m,n}$.
More explicitly, given $\bm{\om}=(\om(z),\hat{\om}(z)) \in
\mathT_{\bm{a}}^* \mathM_{m,n}$ we have
\begin{align}\label{cometricex}
  \eta(\bm{\om})(z)=&\frac{1}{2\pi\bmi}\oint_{|z|=1}\big(\om(p)\eta(da(p))+\hat{\om}(p)\eta(d\hat{a}(p))\big)dp\nn\\
  =&\big(a'(z)[\om(z)+\hat{\om}(z)]_--[\om(z)a'(z)+\hat{\om}(z)\hat{a}'(z)]_-,\nn\\
  &-\hat{a}'(z)[\om(z)+\hat{\om}(z)]_++[\om(z)a'(z)+\hat{\om}(z)\hat{a}'(z)]_+\big),
\end{align}
in which the following formulae have been used:
\begin{align} \label{etad}
  \eta(d\al(p))=(<d\al(p),da(z)>^*,<d\al(p),d\hat{a}(z)>^*),\quad
  \al\in\{a,\hat{a}\}.
\end{align}

\begin{lem}
 The linear map $\eta$ defined in \eqref{linmap} is a bijection.
\end{lem}
\begin{prf}
It follows from \eqref{cometricex} that $\eta$ is surjective, hence
we only need to show that it is invertible.

 Suppose $\bm{\xi}=\eta(\bm{\om})$, where
  \begin{align*}
  \bm{\xi}=&\big(\xi(z),\hat{\xi}(z)\big)=\Big(\sum_{i\leq m}\xi_i z^{2i-2},\sum_{j\geq -n}\hat{\xi}_jz^{2j}\Big)
  \in \mathT_{\bm{a}}\mathM_{m,n}, \\
  \bm{\om}=&\big(\om(z),\hat{\om}(z)\big)=\Big(\sum_{i\leq m}\om_{-i}z^{-2i+1},\sum_{j\geq -n+1}\hat{\om}_{-j}z^{-2j+1}\Big)
  \in \mathT_{\bm{a}}^*\mathM_{m,n}.
  \end{align*}
  Recall $a'(z)-\hat{a}'(z)=\zeta'(z)\ne 0$ in
  \eqref{constraint}, then from \eqref{cometricex} we have
  \begin{align}\label{etain1}
  \om(z)_+=-\Big(\frac{\xi(z)-\hat{\xi}(z)}{a'(z)-\hat{a}'(z)}\Big)_+,\quad
  \hat{\om}(z)_-=\Big(\frac{\xi(z)-\hat{\xi}(z)}{a'(z)-\hat{a}'(z)}\Big)_-.
  \end{align}
  On the other hand, consider
  \begin{align}\label{etain2}
  \xi(z)_+=\big(a'(z)_+(\om(z)+\hat{\om}(z))_-\big)_+
  \end{align}
  where
  \begin{align*}
  a'(z)_+=\ &2mz^{2m-1}+(2m-2)v_mz^{2m-3}+\cdots+2v_2z^1, \\
  (\om(z)+\hat{\om}(z))_-=\
  &\tilde{\om}_{-1}z^{-1}+\tilde{\om}_{-2}z^{-3}+\cdots+\tilde{\om}_{-m}z^{-2m+1}+O(z^{-2m-1}),
  \quad z\to\infty
  \end{align*}
  with $\tilde{\om}_j=\om_j+\hat{\om}_j$.
Introduce the following invertible matrix
\begin{equation}\label{matT}
K_m=\left(
       \begin{array}{ccccc}
       2\,m &  &  &  &  \\
        (2\,m-2)v_m  & 2\,m &  &  &  \\
       (2\,m-4)v_{m-1} & (2\,m-2)v_m & 2\,m &  &  \\
       \vdots & \ddots & \ddots & \ddots &  \\
       2\,v_2 & \cdots & (2\,m-4)v_{m-1}& (2\,m-2)v_m & 2\,m \\
       \end{array}
       \right).
\end{equation}
Equation \eqref{etain2} is  rewritten to
  \begin{equation}\label{etain3}
\left(
  \begin{array}{c}
  \xi_m \\
  \xi_{m-1} \\
  \vdots \\
  \xi_1 \\
  \end{array}
\right)=K_m  \left( \begin{array}{c}
  \tilde{\om}_{-1} \\
  \tilde{\om}_{-2} \\
  \vdots \\
  \tilde{\om}_{-m} \\
  \end{array}
\right).
\end{equation}
Hence $\tilde{\om}_{-1},\dots, \tilde{\om}_{-m}$ can be found from
this linear equation. In combination with \eqref{etain1} one obtains
$(\om(z))_-$ and thus $\om(z)$.

Similarly, from the second equation in \eqref{etain1} together with
\begin{align}\label{etain4}
  \hat{\xi}(z)_-=-\big(\hat{a}'(z)_-(\om(z)+\hat{\om}(z))_+\big)_-
\end{align}
one gets $\hat{\om}(z)$.

Therefore, $\bm\om$ is uniquely determined by $\bm\xi$ via the
relations \eqref{etain1}, \eqref{etain2} and \eqref{etain4}. The
lemma is proved.
\end{prf}

With the help of the bijection $\eta$, we now define a symmetric
bilinear form on the tangent space $\mathT_{\bm{a}}\mathM_{m,n}$ as
\begin{align}\label{metric01}
  <\p_1,\p_2>:=\la
  \eta^{-1}(\p_1),\p_2\ra=<\eta^{-1}(\p_1),\eta^{-1}(\p_2)>^*.
  \end{align}
Recall on $\cM_{m,n}$ the functions
\begin{align}\label{funcpair2}
  \zeta(z)=a(z)-\hat{a}(z),\quad l(z)=a(z)_++\hat{a}(z)_-
\end{align}
satisfy $\zeta'(z)\ne0$ and $l'(z)\ne0$ for $|z|=1$.
\begin{lem}
For any vectors $\p_1, \p_2\in\mathT_{\bm{a}}\mathM_{m,n}$, the
bilinear form \eqref{metric01} is represented as
  \begin{align}
  <\p_1,\p_2>=&-\frac{1}{2\pi\bmi}\oint_{|z|=1}
  \frac{\p_1\zeta(z)\cdot\p_2\zeta(z)}{\zeta'(z)}dz
  \nn\\
  &-\res_{z=\infty}\frac{\p_1l(z)\cdot\p_2l(z)}{l'(z)}dz
  -\res_{z=0}\frac{\p_1l(z)\cdot\p_2l(z)}{l'(z)}dz.
\label{expmetric}
  \end{align}
\end{lem}
\begin{prf}
Suppose $(\om(z),\hat{\om}(z))=\eta^{-1}(\p_1)$. We have
\begin{align}
  \la\eta^{-1}(\p_1),\p_2\ra=&\frac{1}{2\pi\bmi}\oint_{|z|=1}\big(\om(z)\p_2 a(z)+\hat{\om}(z)\p_2\hat{a}(z)\big)dz\nn\\
  =&\frac{1}{2\pi\bmi}\oint_{|z|=1}\big((\om(z)_+-\hat{\om}(z)_-)(\p_2a(z)-\p_2\hat{a}(z))\big)dz\nn\\
  &+\frac{1}{2\pi\bmi}\oint_{|z|=1}(\om(z)_-+\hat{\om}(z)_-)\p_2a(z)dz\nn\\
  &+\frac{1}{2\pi\bmi}\oint_{|z|=1}(\om(z)_++\hat{\om}(z)_+)\p_2\hat{a}(z)dz.
\end{align}
Let $I_1, I_2$ and $I_3$ denote the three integrals on the right
hand side respectively.

First, the formulae \eqref{etain1} now read
\begin{align*}
  \om(z)_+=-\left(\frac{\p_1\zeta(z)}{\zeta'(z)}\right)_+,\quad
  \hat{\om}(z)_-=\left(\frac{\p_1\zeta(z)}{\zeta'(z)}\right)_-,
\end{align*}
hence
\begin{align}
  I_1=&-\frac{1}{2\pi\bmi}\oint_{|z|=1}\frac{\p_1\zeta(z)\cdot\p_2\zeta(z)}{\zeta'(z)}dz.
  \label{int1}
\end{align}

Second, by using the relation \eqref{etain3} we have
\begin{align}
  I_2=&\tilde{\om}_{-1}\p_2v_1+\tilde{\om}_{-2}\p_2v_2+\cdots+\tilde{\om}_{-m}\p_2v_m
\nn \\
=& (\p_1 v_1, \p_1 v_2, \dots, \p_1 v_m)K_m^{-1}\left(
                    \begin{array}{c}
                      \p_2 v_m \\
                      \p_2 v_{m-1} \\
                      \vdots \\
                      \p_2 v_1 \\
                    \end{array}
                    \right).
                    \label{pint3}
\end{align}
Note that ${l'(z)}/{z^{2m-1}}$ is a polynomial in $z^{-2}$, and that
\[
K_m=\left(\frac{l'(z)}{z^{2m-1}}\right)_{z^{-2}\to \Lambda}, \quad
\Lambda=(\delta_{i,j+1})_{m\times m}.
\]
If one writes $1/{l'(z)}=z^{-2m+1}(f_0+f_{-1} z^{-2}+f_{-2}
z^{-4}+\cdots)$, then
\[
K_m^{-1}=\left(\frac{z^{2m-1}}{l'(z)}\right)_{z^{-2}\to
\Lambda}=(f_{-i+j})_{m\times m}.
\]
Thus
\begin{align}
 I_2=&\sum_{i,j=1}^m \p_1 v_i\cdot f_{-i+j}\cdot \p_2 v_{m+1-j}\nn\\
  =& -\res_{z=\infty}\left( \sum_{i=1}^m\p_1v_i z^{2 i-2}\sum_{j=1}^m\p_2v_j z^{2 j-2}
  \cdot z^{-2 m+1}\sum_{k\le0}f_{k}z^{2k}\right) dz \nn\\
  =& -\res_{z=\infty}\frac{\p_1 l(z)\cdot\p_2 l(z)}{ l'(z)} dz.\label{int2}
\end{align}

In the same way one can check
\begin{align}
  I_3=& -\res_{z=0}\frac{\p_1 l(z)\cdot\p_2 l(z)}{ l'(z)} dz.\label{int3}
\end{align}

Finally, we gather \eqref{int1}, \eqref{int2} and \eqref{int3}
together and finish the proof of the lemma.
\end{prf}

\begin{lem}\label{flatc}
 The bilinear form $<\ ,\ >$ in \eqref{expmetric} is a nondegenerate flat metric
 on $\mathM_{m,n}$, with flat coordinates given by \eqref{flatc01}--\eqref{flatc03}.
More exactly,
 \begin{align}
   &<\frac{\p}{\p t^{i_1}},\frac{\p}{\p t^{i_2}}>=-\frac{1}{2}~\dt_{i_1+i_2,0},
   \quad  i_1,i_2\in\mathZ \label{inpdflat01};\\
   &<\frac{\p}{\p h^{j_1}},\frac{\p}{\p h^{j_2}}>=\frac{1}{2m}\dt_{j_1+j_2,m+1},
   \quad j_1,j_2\in\{1,\ldots,m\}; \label{inpdflat02}\\
   &<\frac{\p}{\p \hat{h}^{k_1}},\frac{\p}{\p \hat{h}^{k_2}}>=\frac{1}{2n}\dt_{k_1+k_2,n+1},
   \quad k_1,k_2\in\{1,\ldots,n\} \label{inpdflat03}
 \end{align}
 and any other pairing between these vector fields vanishes.
\end{lem}
\begin{prf}
Recall on $\mathM_{m,n}$ we have assumed that $z\mapsto
w(z)=\zeta(z)^{1/2}$ gives a biholomorphic map from $S^1$ to a
simple curve $\Gm$ around $w=0$. Since $\zeta(z)$ is an even
function, then $w(-z)=-w(z)$. Hence the inverse holomorphic map
$w\mapsto z(w)$ on a neighborhood of $\Gm$ satisfies $z(w)=-z(-w)$.
We apply the Riemann-Hilbert decomposition of this function on the
$w$-plane:
  \begin{align}\label{zw}
   z(w)=f_+(w)+f_-(w),\quad w\in\Gm_1,
  \end{align}
where the functions $f_+(w)$ and $f_-(w)$ are holomorphic inside and
outside the curve $\Gm$ respectively. From the definition
\eqref{flatc01} of $t^i$ it follows that
\[
t^i=\frac{2}{2\pi\bmi}\oint_{\Gm}z(w) w^{-2i}d w,
\]
hence
\begin{align} \label{zw1}
  f_+(w)=\sum_{i\ge1}\frac{t^i}{2} w^{2i-1},\quad |w|\to 0;\\
  f_-(w)=\sum_{i\le0}\frac{t^i}{2} w^{2i-1},\quad
  |w|\to\infty. \label{zw2}
\end{align}

On the other hand, we denote
\begin{align}
\chi(z):=l(z)^{\frac{1}{2m}}\ \mbox{ near }\ \infty, \quad
\hat{\chi}(z):=l(z)^{\frac{1}{2n}}\ \mbox{ near }\ 0.
\end{align}
Similarly as above, the inverse function $z(\chi)$ of $\chi(z)$ is
expanded in the form
\begin{align}
  z(\chi)=\chi-\frac{h^1}{2m}\chi^{-1}-\frac{h^2}{2m}\chi^{-3}-\cdots
  -\frac{h^m}{2m}\chi^{-2m+1}+O(\chi^{-2m-1}), \quad
  |\chi|\to\infty;
\end{align}
and the inverse function $z(\hat{\chi})$ of $\hat{\chi}(z)$ reads
\begin{align}
  z(\hat{\chi})=\frac{\hat{h}^1}{2n}\hat{\chi}^{-1}+\frac{\hat{h}^2}{2n}\hat{\chi}^{-3}+\cdots
  +\frac{\hat{h}^n}{2n}\hat{\chi}^{-2n+1}+O(\hat{\chi}^{-2n-3}),\quad |\hat\chi|\to\infty.
\end{align}

Observe that the functions $\zeta(z)=w(z)^2$ and
$l(z)=\chi(z)^{2m}=\hat{\chi}(z)^{2n}$ are determined by the
variables
\begin{align}\label{thhh}
 \bm{t}\cup\bm{h}\cup\hat{\bm{h}}=\{t^i\mid i\in\mathZ\}\cup\{h^j\mid j=1,\ldots,m\}
 \cup\{\hat{h}^k\mid k=1,\ldots,n\}.
\end{align}
These two functions give the series $a(z)$ and $\hat{a}(z)$ by
\begin{align}\label{reconst}
  a(z)=\zeta(z)_- +l(z),\quad \hat{a}(z)=-\zeta(z)_+ +l(z).
\end{align}
Thus \eqref{thhh} is indeed a system of coordinates on the manifold
$\cM_{m,n}$.

Let us compute the pairings \eqref{inpdflat01}--\eqref{inpdflat03}.
First, due to \eqref{zw}--\eqref{zw2} we have
\[
\frac{\p z(w)}{\p t^i}=\frac{1}{2}w^{2i-1},
\]
hence
\begin{align}\label{formu01}
  \frac{\p w(z)}{\p t^i}=-\frac{1}{2}w(z)^{2i-1}w'(z).
\end{align}
In the same way, for $\chi(z)$ and its inverse function one has
\begin{align}
&  \frac{\p z(\chi)}{\p h^j}=-\frac{1}{2m}\chi^{-2j+1}+O(\chi^{-2m-1}), \quad  |\chi|\to\infty; \nn \\
&   \frac{\p \chi(z)}{\p
h^j}=\frac{1}{2m}\chi(z)^{-2j+1}\chi'(z)+O(z^{-2m-1}), \quad
|z|\to\infty. \label{formu02}
\end{align}
While for $\hat{\chi}(z)$ and its inverse function,
\begin{align}
&   \frac{\p z(\hat{\chi})}{\p \hat{h}^k}=\frac{1}{2n}\hat{\chi}^{-2k+1}
+O(\hat{\chi}^{-2n-1}),\quad |\hat{\chi}|\to\infty; \nn\\
&  \frac{\p \hat{\chi}(z)}{\p
\hat{h}^k}=-\frac{1}{2n}\hat{\chi}(z)^{-2k+1}\hat{\chi}'(z)+O(z^{2n+1}),
\quad |z|\to0.
   \label{formu03}
\end{align}
The formulae \eqref{formu01}--\eqref{formu03} lead to
\begin{align}
\label{dth1}
  \frac{\p\zeta(z)}{\p t^i}=&-w(z)^{2i}w'(z)=-\frac{1}{2}w(z)^{2i-1}\zeta'(z),
  \\
  \label{dlh}
  \frac{\p l(z)}{\p h^j}=&\Big(\chi(z)^{2m-2j}\chi'(z)\Big)_+,\quad
  \frac{\p l(z)}{\p
  \hat{h}^k}=-\Big(\hat{\chi}(z)^{2n-2k}\hat{\chi}'(z)\Big)_-,
  \\
  \frac{\p \zeta(z)}{\p h^j}=&\frac{\p\zeta(z)}{\p\hat{h}^k}=\frac{\p l(z)}{\p t^i}=\Big(\frac{\p l(z)}{\p h^j}\Big)_-
  =\Big(\frac{\p l(z)}{\p \hat{h}^k}\Big)_+=0.
  \label{dth2}
\end{align}
Substituting them into \eqref{expmetric} and by integration by
parts, we have
\begin{align}
  <\frac{\p}{\p t^{i_1}},\frac{\p}{\p t^{i_2}}>=&
  -\frac{1}{2\pi\bmi}\oint_{|z|=1}\frac{w(z)^{2i_1+2i_2-1}w'(z)}{2}dz
  \nn \\
=&-\frac{1}{2\pi\bmi}\oint_{\Gm}\frac{w^{2i_1+2i_2-1}}{2}d w
  =-\frac{1}{2}\dt_{i_1+i_2,0},\nn\\
  <\frac{\p}{\p h^{j_1}},\frac{\p}{\p h^{j_2}}>=&
  -\res_{z=\infty}\frac{\chi(z)^{2m-2j_1-2j_2+1}\chi'(z)}{2m}dz=\frac{1}{2m}\dt_{j_1+j_2,m+1},\nn\\
  <\frac{\p}{\p \hat{h}^{k_1}},\frac{\p}{\p\hat{h}^{k_2}}>
  =&-\res_{z=0}\frac{\hat{\chi}(z)^{2n+1-2k_1-2k_2}\hat{\chi}'(z)}{2n}dz
  =\frac{1}{2n}\dt_{k_1+k_2,n+1}. \nn
\end{align}
Any other pairing between these vectors vanishes. The lemma is
proved.
\end{prf}

\begin{cor}
For the flat metric \eqref{expmetric}, the induced metric on the
cotangent bundle of  $\cM_{m,n}$ is given by \eqref{cometric}.
\end{cor}

According to the identification \eqref{gener}, let us compute
another basis $\{ d t^i\mid i\in\mathbb{Z}\}\cup\{ d h^j\mid 1\le
j\le m\}\cup\{ d \hat{h}^k\mid 1\le k\le n\}$ of $T^*\cM_{m,n}$ that
will be used later. By using the formulae
\eqref{flatc01}--\eqref{flatc03} and \eqref{nacc}, we have
   \begin{align}
  dt^i=&\sum_{r\leq m}\frac{\p t^i}{\p v_r}dv_r+\sum_{r\geq -n}\frac{\p t^i}{\p \hat{v}_r}d\hat{v}_r
\nn\\
=&-\frac{1}{2\pi\bmi}\oint_{|p|=1}w(p)^{-2i-1}
\sum_{r\leq m} \frac{\p \zeta(p)}{\p v_r}dp\cdot (z^{-2r+1},0) \nn\\
  &-\frac{1}{2\pi\bmi}\oint_{|q|=1}w(q)^{-2i-1}
  \sum_{r\geq -n} \frac{\p \zeta(q)}{\p \hat{v}_r}dq\cdot(0,z^{-2r-1}) \nn\\
  =&\Big(-\left(w(z)^{-2i-1}\right)_{\geq-2m+1}, \left(w(z)^{-2i-1}\right)_{\leq
  2n-1}\Big),\label{data01}
\\
  dh^j=&\sum_{r\leq m} \frac{\p h^j}{\p v_r}dv_r
  +\sum_{r\geq -n} \frac{\p h^j}{\p \hat{v}_r}d\hat{v}_r\nn\\
  =&-\res_{p=\infty}~\chi(p)^{2j-2m-1}\sum_{r\leq m} \frac{\p l(p)}{\p v_r}dp\cdot (z^{-2r+1},0)
   \nn\\
   =&\Big(\left(\chi(z)^{2j-2m-1}\right)_{\geq -2m+1},\
   0\Big),\label{data02}
   \\
  d\hat{h}^k=&\sum_{r\leq m} \frac{\p \hat{h}^k}{\p v_r}dv_r+\sum_{r\geq -n}
  \frac{\p \hat{h}^k}{\p \hat{v}_r}d\hat{v}_r\nn\\
  =&\res_{p=0}~\hat{\chi}(p)^{2k-2n-1}\sum_{r\geq -n} \frac{\p l(p)}{\p
  v_r}z^{-2r-2}dp\cdot(0, z^{-2r-1})
  \nn\\
  =&\Big(0\ ,\ \left(\hat{\chi}(z)^{2k-2n-1}\right)_{\leq
  2n-1}\Big).\label{data03}
  \end{align}
Here and below we use notations $\left(\sum_r f_r z^r\right)_{\leq
s}=\sum_{r\le s} f_r z^r$ and $\left(\sum_r f_r z^r\right)_{\geq
s}=\sum_{r\ge s} f_r z^r$.

\subsection{Frobenius algebra}

To construct a Frobenius algebra structure on the tangent spaces of
$\cM_{m,n}$, we start from a multiplication on the cotangent spaces
(recall \eqref{rpgener01} and \eqref{rpgener02}):
\begin{align}\label{comultip}
  d\al(p)\cdot d\beta(q):=\frac{q\beta'(q)}{q^2-p^2}d\al(p)+\frac{p\al'(p)}{p^2-q^2}d\beta(q),
  \quad \al,\beta\in\{a,\hat{a}\}.
\end{align}

\begin{lem}
On $\mathT_{\bm{a}}^* \mathM_{m,n}$ the following two statements
hold true.
\begin{itemize}
\item[(i)] The multiplication \eqref{comultip} is associative and commutative.
 More generally, for $\al_i\in\{a,\hat{a}\}$ one has
  \begin{align}
  d\al_1(p_1)\cdot d\al_2(p_2)\cdot\cdots\cdot d\al_k(p_k)=\sum_{i=1}^{k}\left(\prod_{j\neq
  i}\frac{p_j\al_j'(p_j)}{p_j^2-p_i^2}\right)d\al_i(p_i).
  \end{align}
 \item[(ii)] The bilinear form $<\ ,\ >^*$ defined in \eqref{cometric}
  is invariant with respect to the
 multiplication \eqref{comultip}.
\end{itemize}
\end{lem}
\begin{prf}
It is straightforward to check the first assertion. For the second
one,
\begin{align*}
  &<d\al(p)\cdot d\beta(q),d\gm(r)>^*\nn\\
  =&\frac{q\beta'(q)}{q^2-p^2}\frac{r\gm'(r)}{r^2-p^2}
  +\frac{p\al'(p)}{p^2-q^2}\frac{r\gm'(r)}{r^2-q^2}
  +\frac{p\al'(p)}{p^2-r^2}\frac{q\beta'(q)}{q^2-r^2} \nn\\
  =&<d\al(p),d\beta(q)\cdot d\gm(r)>^*,
\end{align*}
where $\al,\beta,\gm\in\big\{a,\hat{a}\}$. The lemma is proved.
\end{prf}

The multiplication \eqref{comultip} can also be represented in
Laurent series by using \eqref{gener}. In fact, for any
$\bm{\om}_i=(\om_i(z),\hat{\om}_i(z))\in\mathT_{\bm{a}}^*\mathM_{m,n}$
with $i=1, 2$ one can verify
\begin{align}\label{expmult}
  \bm{\om}_1\cdot\bm{\om}_2=&
  \Big(\big[\om_2(z)\big(\om_1(z)a'(z)\big)_+ -\om_2(z)\big(\hat{\om}_1(z)\hat{a}'(z)\big)_-\nn\\
  &-\om_1(z)\big(\om_2(z)a'(z)\big)_-
  -\om_1(z)\big(\hat{\om}_2(z)\hat{a}'(z)\big)_-\big]_{\geq -2m+1},\nn\\
  &\big[\hat{\om}_2(z)\big(\om_1(z)a'(z)\big)_+ +\hat{\om}_2(z)\big(\hat{\om}_1(z)\hat{a}'(z)\big)_+\nn\\
  &+\hat{\om}_1(z)\big(\om_2(z)a'(z)\big)_+
  -\hat{\om}_1(z)\big(\hat{\om}_2(z)\hat{a}'(z)\big)_-\big]_{\leq 2n-1}\Big).
\end{align}
For this multiplication there exists a unity
\begin{equation}
\bm{e}^*:=\left(\frac{z^{-2m+1}}{2m},0\right)=\frac{1}{2m}d h^{1}.
\end{equation}

\begin{prop}
The cotangent space $\mathT_{\bm{a}}^*\mathM_{m,n}$ is a Frobenius
algebra with multiplication defined by \eqref{comultip}, unity
$\bm{e}^*$, and non-degenerate invariant bilinear form
\eqref{cometric}.
\end{prop}

By virtue of the definitions of the bijection $\eta$ in
\eqref{linmap} and the flat metric \eqref{metric01}, we have
\begin{cor}\label{cor-FM}
The tangent space $\mathT_{\bm{a}}\mathM_{m,n}$ is a Frobenius
algebra such that
\begin{itemize}
\item[(i)] The multiplication between vectors $\bm{\xi}_1$ and $\bm{\xi}_2$ is defined by
\begin{align}\label{multip}
  \bm{\xi}_1\cdot\bm{\xi}_2:=\eta\big(\eta^{-1}(\bm{\xi}_1)\cdot\eta^{-1}(\bm{\xi}_2)\big);
\end{align}
\item[(ii)] The unity vector is
\begin{align}\label{unit}
  \bm{e}:=\eta\big(\bm{e}^*\big)=\big(1,1\big)=\frac{\p}{\p
  h^m};
\end{align}
\item[(iii)] The invariant inner product is given by \eqref{expmetric}.
\end{itemize}
\end{cor}

\subsection{The potential and the Euler vector field}

We have introduced a flat metric $<\ ,\ >$ and a Frobenius algebra
structure on the tangent bundle of the manifold $\cM_{m,n}$. Now let
us compute the $3$-tensor
\[
c(\p_1,\p_2,\p_3)= <\p_1\cdot\p_2,\p_3>, \quad \p_1,\p_2,\p_3\in T
\cM_{m,n}.
\]

\begin{lem}\label{correl}
For $i_1, i_2, i_3\in\mathZ$, $j_1,j_2,j_3\in\{1,\ldots,m\}$ and
$k_1,k_2,k_3\in\{1,\ldots,n\}$ it holds that
  \begin{align}
  <\frac{\p}{\p t^{i_1}}\cdot\frac{\p}{\p t^{i_2}},&\frac{\p}{\p t^{i_3}}>=
  -\frac{1}{2\pi \bmi}\oint_{|z|=1}\frac{1}{4}w'(z)\Big[w(z)^{2i_1+2i_2-1}\Pi (w(z)^{2i_3}w'(z))\nn\\
  &+w(z)^{2i_1+2i_3-1}\Pi (w(z)^{2i_2}w'(z))+w(z)^{2i_2+2i_3-1}\Pi (w(z)^{2i_1}w'(z))
  \nn\\&-w(z)^{2(i_1+i_2+i_3-1)}\Pi (w(z)w'(z))\Big]dz \nn\\
  &-\frac{1}{2\pi\bmi}\oint_{|z|=1}\frac{1}{4}l'(z)w(z)^{2(i_1+i_2+i_3-1)}w'(z)dz
 \end{align}
with $\Pi (f(z))=f(z)_+-f(z)_-$, and
 \begin{align}
  <\frac{\p}{\p t^{i_1}}\cdot\frac{\p}{\p t^{i_2}},&\frac{\p}{\p h^{j_3}}>
  =-\frac{1}{2\pi\bmi}\oint_{|z|=1}\frac{1}{2}\big(w(z)^{2i_1+2i_2-1}w'(z)\big)_-
  \big(\chi(z)^{2m-2j_3}\chi'(z)\big)_+ d z,
  \\
  <\frac{\p}{\p t^{i_1}}\cdot\frac{\p}{\p t^{i_2}},&\frac{\p}{\p \hat{h}^{k_3}}>
  =\frac{1}{2\pi\bmi}\oint_{|z|=1}\frac{1}{2}\big(w(z)^{2i_1+2i_2-1}w'(z)\big)_+
  \big(\hat{\chi}(z)^{2n-2k_3}\hat{\chi}'(z)\big)_- d z,
  \\
  <\frac{\p}{\p h^{j_1}}\cdot\frac{\p}{\p h^{j_2}},&\frac{\p}{\p t^{i_3}}>
  =-\frac{1}{2\pi\bmi}\oint_{|z|=1}\frac{1}{2m}
  \big(\chi(z)^{2m-2j_1-2j_2+1}\chi'(z)\big)_+\big(w(z)^{2i_3}w'(z)\big)_- d z,
  \\
    <\frac{\p}{\p h^{j_1}}\cdot\frac{\p}{\p h^{j_2}},&\frac{\p}{\p \hat{h}^{k_3}}>
  =-\frac{1}{2\pi\bmi}\oint_{|z|=1}\frac{1}{2m}\big(\chi(z)^{2m-2j_1-2j_2+1}\chi'(z)\big)_+
  \big(\hat{\chi}(z)^{2n-2k_3}\hat{\chi}'(z)\big)_-dz,
  \\
  <\frac{\p}{\p \hat{h}^{k_1}}\cdot\frac{\p}{\p \hat{h}^{k_2}},&\frac{\p}{\p t^{i_3}}>
  =-\frac{1}{2\pi\bmi}\oint_{|z|=1}\frac{1}{2n}
  \big(\hat{\chi}(z)^{2n-2k_1-2k_2+1}\hat{\chi}'(z)\big)_-\big(w(z)^{2i_3}w'(z)\big)_+dz,
  \\
  <\frac{\p}{\p \hat{h}^{k_1}}\cdot\frac{\p}{\p \hat{h}^{k_2}},&\frac{\p}{\p h^{j_3}}>
  =-\frac{1}{2\pi\bmi}\oint_{|z|=1}\frac{1}{2n}\big(\hat{\chi}(z)^{2n-2k_1-2k_2+1}
  \hat{\chi}'(z)\big)_-\big(\chi(z)^{2m-2j_3}\chi'(z)\big)_+dz,
\\
  <\frac{\p}{\p h^{j_1}}\cdot\frac{\p}{\p h^{j_2}},&\frac{\p}{\p h^{j_3}}>
  =-\frac{1}{2\pi\bmi}\oint_{|z|=1}\frac{1}{2m^2}\big(w(z)w'(z)\big)_-
  \big(\chi(z)^{2m-2j_1-2j_2-2j_3+2}\chi'(z)\big)_+dz
  \nn\\
  &-\res_{z=\infty}\frac{\big(\chi(z)^{2m-2j_1}\chi'(z)\big)_+\big(\chi(z)^{2m-2j_2}
  \chi'(z)\big)_+\big(\chi(z)^{2m-2j_3}\chi'(z)\big)_+}{l'(z)}dz, \\
  <\frac{\p}{\p \hat{h}^{k_1}}\cdot\frac{\p}{\p \hat{h}^{k_2}},&\frac{\p}{\p \hat{h}^{k_3}}>
  =\frac{1}{2\pi\bmi}\oint_{|z|=1}\frac{1}{2n^2}\big(w(z)w'(z)\big)_+
  \big(\hat{\chi}(z)^{2n-2k_1-2k_2-2k_3+2}\hat{\chi}'(z)\big)_-dz \nn\\
  &+\res_{z=0}\frac{\big(\hat{\chi}(z)^{2m-2k_1}\hat{\chi}'(z)\big)_-\big(\hat{\chi}(z)^{2m-2k_2}\hat{\chi}'(z)\big)_-
  \big(\hat{\chi}(z)^{2m-2k_3}\hat{\chi}'(z)\big)_-}{l'(z)}dz,
  \\
<\frac{\p}{\p t^{i_1}}\cdot \frac{\p}{\p h^{j_2}},&\frac{\p}{\p
\hat{h}^{k_3}}>=0.
  \end{align}
\end{lem}
\begin{prf}
  Denote $\eta_{ u v}=<\p_u,\p_v>$ for $u,v\in\bm{t}\cup\bm{h}\cup\hat{\bm{h}}$, then
  we have
  \begin{align}
    <\p_u\cdot \p_v,\p_w>
    =\eta_{u\tilde{u}}\eta_{v\tilde{v}}\eta_{w\tilde{w}}<d\tilde{u}\cdot d\tilde{v},d\tilde{w}>^*
    =\eta_{u\tilde{u}}\eta_{v\tilde{v}}\eta_{w\tilde{w}}\la d\tilde{u}
    \cdot d\tilde{v},\eta(d\tilde{w})\ra.
  \end{align}
  Substitute the formulae \eqref{data01}--\eqref{data03} into
  the right hand side and use \eqref{expmult} and \eqref{cometricex},
  then
  the lemma follows from a lengthy but straightforward computation.
\end{prf}

\begin{lem}\label{lem-finF}
There exists a function $F_{m,n}$ depending rationally on the
variables $\bm{h}\cup\hat{\bm{h}}$ such that
  \begin{align}\label{poten03}
    \frac{\p^3 F_{m,n}}{\p u\, \p v\, \p w}
    =-\big(\res_{z=\infty}+\res_{z=0}\big)\frac{\p_{u} l(z)\cdot\p_{v}
    l(z)\cdot\p_{w} l(z)}{l'(z)}dz
  \end{align}
  for any $u,v, w\in\bm{h}\cup\hat{\bm{h}}$.
\end{lem}

This lemma will be proved in the Appendix. Now we have the following
result.
\begin{prop}\label{prp-F}
On $\mathM_{m,n}$ the following function
     \begin{align}\label{poten2}
     \mathcal{F}_{m,n}=&\left(\frac{1}{2\pi\bmi}\right)^2
     \oint\oint_{|z_1|<|z_2|}\left(\frac{1}{2}\zeta(z_1)\zeta(z_2)-\zeta(z_1)l(z_2)
     +l(z_1)\zeta(z_2)\right)\times \nn \\
     & \qquad\qquad\qquad \times\log\left(\frac{z_2-z_1}{z_2}\right)dz_1dz_2+F_{m,n}
  \end{align}
  satisfies
\begin{equation}
c(\p_u,\p_v,\p_w)=\frac{\p^3 \mathcal{F}_{m,n}}{\p u\,\p v\,\p
w},\quad u,v,w\in\bm{t}\cup\bm{h}\cup\hat{\bm{h}}.
\end{equation}
This implies that
  $\nabla_{s}c(\p_u,\p_v,\p_w)$ is a symmetric $4$-tensor, where $\nabla$ stands
  for the Levi-Civita connection of the metric $<\ ,\ >$.
\end{prop}

\begin{prf}
  The proposition can be checked with the help of
  the formulae \eqref{dth1}--\eqref{dth2} and
  by integration by parts as follows.

Denote
\begin{equation}\label{}
Y_i(z)=\left\{\begin{array}{cl}
         \dfrac{1}{i+1}\zeta(z)^{(i+1)/2}=\dfrac{1}{i+1}w(z)^{i+1},\quad  & i\in\mathbb{Z}\setminus\{-1\}; \\ \\
         \log\zeta(z)^{1/2}=\log w(z), & i=-1.
       \end{array}\right.
\end{equation}
Clearly, all $Y_i'(z)$ are holomorphic at $|z|=1$, and
\begin{equation}\label{}
\frac{1}{2\pi \bmi}\oint_{|z|=1}Y_i'(z) d z=\delta_{i,-1}.
\end{equation}

It follows from \eqref{dth1} that
\begin{align}
&\frac{\p\zeta(z)}{\p t^i}=-\frac{1}{2}\zeta(z)^{i-\frac{1}{2}}\zeta'(z)=-Y_{2i}'(z), \\
&\frac{\p^2\zeta(z)}{\p t^{i_1}\p
t^{i_2}}=\p_z\left(\frac{1}{4}\zeta(z)^{i_1+i_2-1}\zeta'(z)\right)=\frac{1}{2}Y_{2i_1+2i_2-1}''(z),\\
&\frac{\p^3\zeta(z)}{\p t^{i_1}\p t^{i_2}\p
t^{i_3}}=-\p_z^2\left(\frac{1}{8}\zeta(z)^{i_1+i_2+i_3-\frac{3}{2}}\zeta'(z)\right)
=-\frac{1}{4}Y_{2i_1+2i_2+2i_3-2}'''(z).
\end{align}
Then we have
\begin{align}
&\frac{\p^3\mathcal{F}_{m,n} }{\p t^{i_1}\,\p t^{i_2}\,\p
t^{i_3}}\nn\\
 =&\frac{1}{(2\pi \bmi)^2}\oint\oint_{|z_1|<|z_2|}
 \bigg(-\frac{1}{8}Y_{2i_1+2i_2+2i_3-2}'''(z_1)\zeta(z_2)-
 \frac{1}{8}\zeta(z_1)Y_{2i_1+2i_2+2i_3-2}'''(z_2) \nn\\
 &\qquad -\frac{1}{4}\sum_{\mathrm{c.p.}(i_1,i_2,i_3)}\left(
 Y_{2i_1+2i_2-1}''(z_1)Y_{2i_3}'(z_2)+Y_{2i_3}'(z_1)Y_{2i_1+2i_2-1}''(z_2)
 \right) \nn\\
 &\qquad +\frac{1}{4}Y_{2i_1+2i_2+2i_3-2}'''(z_1)l(z_2)-\frac{1}{4}l(z_1)Y_{2i_1+2i_2+2i_3-2}'''(z_2)\bigg)
 \log\left(\frac{z_2-z_1}{z_2}\right) d z_1 d z_2 \nn\\
=&\frac{1}{(2\pi \bmi)^2}\oint\oint_{|z_1|<|z_2|}
 \bigg(-\frac{1}{8}Y_{2i_1+2i_2+2i_3-2}''(z_1)\zeta(z_2)+
 \frac{1}{8}\zeta(z_1)Y_{2i_1+2i_2+2i_3-2}''(z_2) \nn\\
 &\qquad -\frac{1}{4}\sum_{\mathrm{c.p.}(i_1,i_2,i_3)}\left(
 Y_{2i_1+2i_2-1}'(z_1)Y_{2i_3}'(z_2)-Y_{2i_3}'(z_1)Y_{2i_1+2i_2-1}'(z_2)
 \right) \nn\\
 &\qquad +\frac{1}{4}Y_{2i_1+2i_2+2i_3-2}''(z_1)l(z_2)
 +\frac{1}{4}l(z_1)Y_{2i_1+2i_2+2i_3-2}''(z_2)\bigg)\frac{ d z_1 d z_2}{z_2-z_1}
  \nn\\
 &+\frac{1}{(2\pi \bmi)^2}\oint\oint_{|z_1|<|z_2|}
 \bigg(-\frac{1}{8}\zeta(z_1)Y_{2i_1+2i_2+2i_3-2}''(z_2)
 -\frac{1}{4}\sum_{\mathrm{c.p.}(i_1,i_2,i_3)}Y_{2i_3}'(z_1)Y_{2i_1+2i_2-1}'(z_2)
  \nn\\
 &\qquad -\frac{1}{4}l(z_1)Y_{2i_1+2i_2+2i_3-2}''(z_2)\bigg)\frac{ d z_1 d z_2}{z_2} \nn\\
=&\frac{1}{2\pi \bmi}\oint
 \bigg(-\frac{1}{8}Y_{2i_1+2i_2+2i_3-2}''(z_2)_-\zeta(z_2)+
 \frac{1}{8}\zeta(z_2)_{-}Y_{2i_1+2i_2+2i_3-2}''(z_2) \nn\\
 &\qquad -\frac{1}{4}\sum_{\mathrm{c.p.}(i_1,i_2,i_3)}\left(
 Y_{2i_1+2i_2-1}'(z_2)_{-}Y_{2i_3}'(z_2)-Y_{2i_3}'(z_2)_{-}Y_{2i_1+2i_2}'(z_2)
 \right) \nn\\
 &\qquad +\frac{1}{4}Y_{2i_1+2i_2+2i_3-2}''(z_2)_{-}l(z_2)
 +\frac{1}{4}l(z_2)_{-}Y_{2i_1+2i_2+2i_3-2}''(z_2)\bigg)
  d z_2 + 0\nn\\
=&\frac{1}{2\pi \bmi}\oint
 \bigg(\frac{1}{4}Y_{2i_1+2i_2+2i_3-2}'(z)\left(\frac{1}{2}\zeta'(z)_{+}-
 \frac{1}{2}\zeta'(z)_{-}-l'(z)_{+}-l'(z)_{-} \right) \nn\\
 &\qquad -\frac{1}{4}\sum_{\mathrm{c.p.}(i_1,i_2,i_3)}
 Y_{2i_1+2i_2-1}'(z)(Y_{2i_3}'(z)_{+}-Y_{2i_3}'(z)_{-}) \bigg)
  d z \nn\\
=&\frac{1}{2\pi \bmi}\oint
 \bigg(\frac{1}{4}Y_{2i_1+2i_2+2i_3-2}'(z)\left(\frac{1}{2}\Pi\zeta'(z)-l'(z) \right) \nn\\
 &\qquad -\frac{1}{4}\sum_{\mathrm{c.p.}(i_1,i_2,i_3)}
 Y_{2i_1+2i_2-1}'(z) \Pi\,Y_{2i_3}'(z) \bigg)
  d z \nn\\
=&<\p_{t^{i_1}}\cdot \p_{t^{i_2}},\p_{t^{i_3}}>,
\end{align}
where ``c.p.'' stands for ``cyclic permutation''.

In the same way, by using \eqref{dlh} one has
\begin{align}
&\frac{\p^3\mathcal{F}_{m,n} }{\p t^{i_1}\,\p t^{i_2}\,\p
\hat{h}^{k_3}}\nn\\
 =&\frac{1}{(2\pi \bmi)^2}\oint\oint_{|z_1|<|z_2|}
 \bigg( \frac{1}{2}Y_{2i_1+2i_2-1}''(z_1)(\hat{\chi}(z_2)^{2n-2k_3}\hat{\chi}'(z_2))_{-}
 \nn\\
 &\qquad
 -\frac{1}{2}(\hat{\chi}(z_1)^{2n-2k_3}\hat{\chi}'(z_1))_{-}Y_{2i_1+2i_2-1}''(z_2)\bigg)
 \log\left(\frac{z_2-z_1}{z_2}\right) d z_1 d z_2
 \nn\\
 =&\frac{1}{(2\pi \bmi)^2}\oint\oint_{|z_1|<|z_2|}
 \bigg( \frac{1}{2}Y_{2i_1+2i_2-1}'(z_1)(\hat{\chi}(z_2)^{2n-2k_3}\hat{\chi}'(z_2))_{-}
 \nn\\
 &\qquad
 +\frac{1}{2}(\hat{\chi}(z_1)^{2n-2k_3}\hat{\chi}'(z_1))_{-}Y_{2i_1+2i_2-1}'(z_2)\bigg)
 \frac{  d z_1 d z_2 }{z_2-z_1}\nn\\
 &-\frac{1}{(2\pi \bmi)^2}\oint\oint_{|z_1|<|z_2|}
\frac{1}{2}(\hat{\chi}(z_1)^{2n-2k_3}\hat{\chi}'(z_1))_{-}Y_{2i_1+2i_2-1}'(z_2)
 \frac{ d z_1 d z_2}{z_2} \nn\\
 =&\frac{1}{2\pi \bmi}\oint
 \bigg( 0+\frac{1}{2}(\hat{\chi}(z_2)^{2n-2k_3}\hat{\chi}'(z_2))_{-}Y_{2i_1+2i_2-1}'(z_2)_+\bigg)
  d z_2+0
 \nn\\
 =&\frac{1}{2\pi \bmi}\oint
 \frac{1}{2}Y_{2i_1+2i_2-1}'(z)_{+}(\hat{\chi}(z)^{2n-2k_3}\hat{\chi}'(z))_{-} d
 z
\nn\\
=&<\p_{t^{i_1}}\cdot \p_{t^{i_2}},\p_{\hat{h}^{k_3}}>.
\end{align}
The other cases are similar.
 Thus the proposition is proved.
\end{prf}

To show that  $\cM_{m,n}$ is a Frobenius manifold, we still need to
fix a Euler vector field and show the quasi-homogeneity of
$\mathcal{F}_{m,n}$.

We assign a degree to each of the variables as
 \begin{align}\label{degree}
   \deg\,t^i=\frac{m(1-2i)+1}{2m},\quad \deg\,h^j=\frac{j}{m},\quad
   \deg\,\hat{h}^k=\frac{2k-1}{2n}+\frac{1}{2m},
 \end{align}
 and let
  \begin{align}\label{euler2}
    \mathcal{E}_{m,n}=\sum_{i\in\mathZ}\frac{m(1-2i)+1}{2m}t^i
    \frac{\p}{\p t^i}+\sum_{j=1}^m\frac{j}{m}h^j\frac{\p}{\p h^j}
     +\sum_{k=1}^n\Big(\frac{2k-1}{2n}+\frac{1}{2m}\Big)\hat{h}^k\frac{\p}{\p\hat{h}^k}.
  \end{align}
\begin{lem} \label{EF}
The function $\mathcal{F}_{m,n}$ is quasi-homogeneous with respect
to the degrees \eqref{degree}. More precisely, we have (cf.
\eqref{WDVV2})
\begin{equation}
Lie_{ \mathcal{E}_{m,n}}\mathcal{F}_{m,n}=(3-d_m)\mathcal{F}_{m,n},
\quad d_m=1-\frac1{m}.
\end{equation}
\end{lem}
\begin{prf}
In addition to \eqref{degree} we assume $\deg\,z=1/2m$. Note that
both functions $\zeta(z)$ and $l(z)$ are homogeneous of degree $1$,
hence $\mathcal{F}_{m,n} $ has degree $2+1/m$. The lemma is proved.
\end{prf}

Up to now,  we have shown that $\cM_{m,n}$ is a Frobenius manifold,
on which the multiplication is defined by \eqref{multip} with unity
$\bm{e}=\p/\p h^m$ and invariant flat metric \eqref{expmetric}, the
Euler vector field is $\mathcal{E}_{m,n}$, and the potential is
$\mathcal{F}_{m,n}$.

\subsection{Semisimplicity}

The final step in proving Theorem~\ref{main} is to show the
semisimplicity of $\cM_{m,n}$.

Let
\begin{equation}\label{}
d\mu(z)=\frac{d a(z)}{a'(z)}-\frac{d \hat{a}(z)}{\hat{a}'(z)}, \quad
z\in S^1.
\end{equation}
This is a generating function for a basis of the cotangent space
$T^*_{\bm{a}}\mathM_{m,n}$ by taking the Riemann-Hilbert
decomposition with respect to $S^1$. According to \eqref{cometric}
and \eqref{comultip}, it is easy to show the following lemma.
\begin{lem}\label{lem-dmu}
The following formulae hold true:
\begin{align} \label{dmupai}
  &<d\mu(p),d\mu(q)>^*=-\frac{\zeta'(p)}{a'(p)\hat{a}'(p)}\dt_0(p-q), \\
  &d\mu(p)\cdot
  d\mu(q)=\dt_0(p-q)\,d\mu(p), \label{dmupro}
\end{align}
where $\dt_0(p-q)=\sum_{k\in\mathZ}\left({p^{2k}}/{q^{2k+1}}\right)$
satisfies
\begin{align}
\frac{1}{2\pi\bmi}\oint_{|q|=1}f(q)\dt_0(p-q)\,d
q=\left\{\begin{array}{cc}
                                                          f(p), & f(-q)=f(q); \\
                                                          0, &
                                                          f(-q)=-f(q).
                                                        \end{array}\right.
\end{align}
\end{lem}

The formula \eqref{dmupro} implies that no nilpotent elements exist
in the Frobenius algebra $T^*_{\bm{a}}\cM_{m,n}$, hence the
Frobenius algebra $T^*_{\bm{a}}\cM_{m,n}$ is semisimple and so is
$T_{\bm{a}}\cM_{m,n}$. Thus we arrive at the following result.
\begin{prop}\label{prp-ss}
The Frobenius manifold $\cM_{m,n}$ is semisimple.
\end{prop}

\begin{prfn}{Theorem~\ref{main}}
The theorem follows from a combination of Lemmas~\ref{flatc}
and~\ref{EF}, Corollary~\ref{cor-FM}, Propositions~\ref{prp-F} and
\ref{prp-ss}.
\end{prfn}

We want to find canonical coordinates on a subset
$\mathM_{m,n}^{s}\subset \mathM_{m,n}$ consisting of generic points
$(a(z),\hat{a}(z))$ that satisfy the following conditions
\begin{description}
\item[(S1)] For $z\in\mathS^1$,
\begin{align}\label{canocond}
a'(z)\hat{a}''(z)-\hat{a}'(z)a''(z)\neq 0;
\end{align}
\item[(S2)]
The function $\sg(z)=\big({a'(z)}/{\zeta'(z)}\big)^{1/2}$ is
holomorphic and injective on $S^1$ such that a smooth simple closed
curve is defined by
\begin{align}\label{curve}
  \Sigma:=\left\{\sg=\left(\frac{a'(z)}{\zeta'(z)}\right)^{\frac{1}{2}} \mid z\in S^1\right\}.
\end{align}
\end{description}

 Denote the inverse function of
$\sg(z)$ by $z=z(\sg): \Sigma\to S^1$, then
\begin{align}\label{canocoord}
  \big[\sg^2\hat{a}'(z)+(1-\sg^2)a'(z)\big]_{z=z(\sg)}=0,\quad \sg\in
  \Sigma.
\end{align}
Let
\begin{align}\label{canocoor}
  u_{\sg}:=\big[\sg^2\hat{a}(z)+(1-\sg^2)a(z)\big]_{z=z(\sg)},\quad \sg\in
  \Sigma.
\end{align}
One sees that the exact $1$-form $d u_\sg$ is just a normalization
of the generating function $d\mu(z)$, more exactly,
  \begin{align}
  du_{\sg}=\left(\frac{a'(z)}{\zeta'(z)}d\hat{a}(z)-\frac{\hat{a}'(z)}{\zeta'(z)}da(z) \right)_{z=z(\sg)}
  =\left(-\frac{a'(z)\hat{a}'(z)}{\zeta'(z)}d\mu(z)\right)_{z=z(\sg)}.
 \end{align}
Thus from Lemma~\ref{lem-dmu}  it follows that $u_\sg$ is a
canonical coordinate on $\mathM_{m,n}^{s}$ (in the sense of
\cite{CDM}). In fact, it can be checked that
$\mathcal{E}_{m,n}(u_\sg)=u_\sg$ by using the formulae \eqref{Eaah}
below.

\subsection{The intersection form}

Define the intersection form on the cotangent space
$\mathT^*_{\bm{a}}\mathM_{m,n}$ as (cf. \eqref{inters})
  \begin{align}\label{iinters}
  \big(d\al(p),d\beta(q)\big)^*:=i_{\mathcal{E}_{m,n}}\big(d\al(p)\cdot d\beta(q)\big),
  \quad \al,\beta\in\big\{a,\hat{a}\big\}.
  \end{align}

\begin{lem}
For $ \al,\beta\in\big\{a,\hat{a}\big\}$ it holds that
  \begin{align}\label{iinters2}
  \big(d\al(p),d\beta(q)\big)^*=
  \frac{q\beta'(q)}{q^2-p^2}\al(p)+\frac{p\al'(p)}{p^2-q^2}\beta(q).
  \end{align}
\end{lem}
\begin{prf}
In consideration of the degree of coordinates in \eqref{degree}, one
can express the Euler vector field \eqref{euler2} in the coordinates
$\{v_i\}_{i=m}^{-\infty}\cup\{\hat{v}_j\}_{j=-n}^{+\infty}$ as
follows:
\begin{align}
  \mathcal{E}_{m,n}=\sum_{i\leq m}\frac{m+1-i}{m}v_i\frac{\p}{\p v_i}
   +\sum_{j\geq -n}\frac{m-j}{m}\hat{v}_j\frac{\p}{\p \hat{v}_j}.
   \label{euler02}
\end{align}
Actually this formula can also be checked by letting \eqref{euler02}
act on the flat coordinates \eqref{flatc01}--\eqref{flatc03}.

It is easy to see
\begin{equation}\label{Eaah}
\mathcal{E}_{m,n}(\al(z))=\al(z)-\frac{z}{2m}\al'(z), \quad
\al,\beta\in\big\{a,\hat{a}\big\}.
\end{equation}
Substituting \eqref{comultip} into \eqref{iinters} we have
\begin{align}
  \big(d\al(p),d\beta(q)\big)^*
  =&\left\la\mathcal{E}_{m,n},
   \frac{q\beta'(q)}{q^2-p^2}da(p)+\frac{pa'(p)}{p^2-q^2}d\beta(q)\right\ra
   \nn\\
   =&\frac{q\beta'(q))}{q^2-p^2}\left(\al(p)-\frac{p}{2m} \al'(p)\right)
   +\frac{p \al'(p)}{p^2-q^2}\left(\beta(q)-\frac{q}{2m}\beta'(q)\right)
   \nn\\
   =&\frac{q\beta'(q)}{q^2-p^2}\al(p)+\frac{pa'(p)}{p^2-q^2}\beta(q).\nn
  \end{align}
The lemma is proved.
\end{prf}

Similar to the case of the flat metric \eqref{cometric}, the
intersection form \eqref{iinters} induces another metric on the
tangent space $\mathT_{\bm{a}}\mathM_{m,n}$.

In fact, there is a linear map
\begin{align}\label{linmapg}
g:\ \mathT_{\bm{a}}^*\mathM_{m,n}\rightarrow
\mathT_{\bm{a}}\mathM_{m,n}
\end{align}
such that
\[
\la \bm{\om}_1,g(\bm{\om}_2)\ra=(\bm{\om}_1,\bm{\om}_2)^*
\]
for any $\bm{\om}_1,\bm{\om}_2\in\mathT_{\bm{a}}^*\mathM_{m,n}$.
Given  any
$\bm{\om}=\big(\om(z),\hat{\om}(z)\big)\in\mathT_{\bm{a}}^*\mathM_{m,n}$,
the formulae (cf. \eqref{etad} and \eqref{cometricex})
\begin{align}
  g(d\al(p))=\Big(\big(d\al(p),da(z)\big)^*,\big(d\al(p),d\hat{a}(z)\big)^*\Big),
  \quad \al\in\{a,\hat{a}\}
\end{align}
lead to
\begin{align}
  g(\bm{\om})(z)=&\Big(a'(z)\big(a(z)\om(z)+\hat{a}(z)\hat{\om}(z)\big)_--a(z)\big(a'(z)\om(z)+\hat{a}'(z)\hat{\om}(z)\big)_-,\nn\\
  &-\hat{a}'(z)\big(a(z)\om(z)+\hat{a}(z)\hat{\om}(z)\big)_+
  +\hat{a}(z)\big(a'(z)\om(z)+\hat{a}'(z)\hat{\om}(z)\big)_+\Big).
\end{align}
This shows that the map $g$ is surjective. On the other hand,
suppose
$g(\bm{\om})=\big(\xi(z),\hat{\xi}(z)\big)\in\mathT_{\bm{a}}\mathM_{m,n}$,
then
\begin{align}
  &\om(z)=\Big[\frac{1}{a(z)}\Big(\frac{\hat{a}(z)\xi(z)-a(z)\hat{\xi}(z)}{a(z)\hat{a}'(z)-a'(z)\hat{a}(z)}\Big)_+
  \Big]_{\geq -2m+1},\label{invinter01}\\
  &\hat{\om}(z)=-\Big[\frac{1}{\hat{a}(z)}\Big(\frac{\hat{a}(z)\xi(z)-a(z)\hat{\xi}(z)}{a(z)\hat{a}'(z)-a'(z)\hat{a}(z)}\Big)_-
  \Big]_{\leq 2n-1}. \label{invinter02}
\end{align}
Thus $g$ is injective. Therefore, the linear map $g$ defined in
\eqref{linmapg} is a bijection.

Now we define a bilinear form on $T_{\bm{a}}\mathM_{m,n}$ as
\begin{align}\label{metric02}
  \big(\p_1,\p_2\big):=\la g^{-1}(\p_1),\p_2\ra=\big(g^{-1}(\p_1),g^{-1}(\p_2)\big)^*
\end{align}

\begin{prop}
The Frobenius manifold $\cM_{m,n}$ has an intersection form
\eqref{iinters}, which induces a flat metric as
\begin{align}\label{intsf}
  &\big(\p_1,\p_2\big)
  =-\frac{1}{2\pi\bmi}\oint_{|z|=1}\frac{\p_1\log \big(a(z)/\hat{a}(z)\big)\cdot
  \p_2\log \big(a(z)/\hat{a}(z)\big)}
  {\p_z\log \big(a(z)/\hat{a}(z)\big)}dz
\end{align}
with arbitrary vectors $\p_1, \p_2\in T_{\bm{a}}\mathM_{m,n}$.
\end{prop}
\begin{prf}
The equality \eqref{intsf} follows from the formulae
\eqref{invinter01}--\eqref{metric02}. The flatness of this metric
can be verified with the same method as used in Lemma~\ref{flatc}.
\end{prf}

\section{Principal two-component BKP hierarchy}
\label{sec-bkp}

We are to write down the principal hierarchy for the Frobenius
manifold $\mathM_{m,n}$. First let us consider the bi-Hamiltonian
structure associated to this Frobenius manifold, i.e., two
compatible Hamiltonian structures corresponding to the flat metrics
\eqref{cometric} and \eqref{iinters}.

Let $\mathL \mathM_{m,n}$ be the loop space of smooth maps from
$\mathS^1$ to the manifold $\mathM_{m,n}$. In this space a point is
expressed as $\bm{a}=\big(a(z,x),\hat{a}(z,x)\big)$ in \eqref{aah}
with the coefficients being smooth functions of $x\in\mathS^1$. The
tangent space and cotangent space at any point of $\mathL
\mathM_{m,n}$ are composed of smooth maps from $\mathS^1$ to
$z^{2m-2}\mathH^-\times z^{-2n}\mathH^+$ and to
$z^{-2m+1}\mathH^+\times z^{2n-1}\mathH^-$ respectively. The pairing
between a cotangent vector
$\bm{\om}=\big(\om(z,x),\hat{\om}(z,x)\big)$ and a tangent vector
$\bm{\xi}=\big(\xi(z,x),\hat{\xi}(z,x)\big)$ is (cf.
\eqref{dualpair})
\begin{align}
  \la\bm{\om},\bm{\xi}\ra=\frac{1}{2\pi\bmi}\oint_{\mathS
  ^1}\oint_{|z|=1}
  \big[\om(z,x)\xi(z,x)+\hat{\om}(z,x)\hat{\xi}(z,x)\big]d z\, d x.
\end{align}
Without confusion we will simply write $(\al(z,x), \hat{\al}(z,x))$
as $(\al(z), \hat{\al}(z))$ below.

There uniquely exist
\begin{align}\label{rela}
\lm(z)=z+\cdots=a(z)^{\frac{1}{2m}} \hbox{ near } \infty, \quad
\hat{\lm}(z)=\frac{\hat{h}^1}{2 n}
z^{-1}+\cdots=\hat{a}(z)^{\frac{1}{2n}} \hbox{ near }  0.
\end{align}
On the loop space $\mathL \mathM_{m,n}$ a hierarchy of evolutionary
equations is defined as
\begin{align}
  &\frac{\p\lm(z)}{\p s_k}=\{(\lm(z)^k)_+,\lm(z)\},\quad
  \frac{\p\hat{\lm}(z)}{\p s_k}=\{(\lm(z)^k)_+,\hat{\lm}(z)\},\label{disphir01}\\
  &\frac{\p\lm(z)}{\p
  \hat{s}_k}=\{-(\hat{\lm}(z)^k)_-,\lm(z)\},\quad
  \frac{\p\hat{\lm}(z)}{\p\hat{s}_k}=\{-(\hat{\lm}(z)^k)_-,\hat{\lm}(z)\},
  \label{disphir02}
\end{align}
where $k\in\mathZ_+^{odd}$ and the Lie bracket $\big\{f,g\big\}:=\p
f/ \p z \cdot \p g/\p x-\p g/\p z\cdot\p f/\p x$. This hierarchy is
called the dispersionless two-component BKP hierarchy.

\begin{rem}
The system of Lax equations \eqref{disphir01}, \eqref{disphir02} was
first written down by  Takasaki \cite{Takasaki} as the universal
hierarchy underlying the D-type topological Landau-Ginzburg models.
The name presently used is after the bilinear equations (whose
dispersionless limit is equivalent to \eqref{disphir01},
\eqref{disphir02}) introduced by Date, Jimbo, Kashiwarw and Miwa
\cite{DJKM} considering the neutral free-fermion realization of the
Lie algebra with infinite Dynkin diagram $D_\infty$, see also
\cite{JM, Kac, LWZ, Shiota}.
\end{rem}

A map $\mathcal{P}$ from the cotangent space to the tangent space of
$\mathcal{L} \mathM_{m,n}$ is a Poisson structure if it defines a
Poisson bracket between local functionals on $\mathL \mathM_{m,n}$
as
\begin{align}
  \big\{F,H\big\}_\mathcal{P}=\la d F,\mathcal{P}(d H)\ra.
\end{align}
Here $d F$ denotes the gradient of the functional $F$, namely, the
covector on $\mathL \mathM_{m,n}$ determined by $\dt
F=\la\dt\bm{a},dF\ra$; and $d H$ means the same.

\begin{prop}[\cite{Wu}]
The space  $\mathcal{L} \mathM_{m,n}$ carries the following two
compatible Poisson structures:
\begin{align}
  \mathcal{P}_1&(\om(z),\hat{\om}(z)) \nn\\
  =&\Big(-(\{a(z),\om(z)\}+\{\hat{a}(z),\hat{\om}(z)\})_-
  +\{a(z),(\om(z)+\hat{\om}(z))_-\}, \nn
\\
  &\quad (\{a(z),\om(z)\}+\{\hat{a}(z),\hat{\om}(z)\})_+
  -\{\hat{a}(z),(\om(z)+\hat{\om}(z))_+\}\Big),
  \label{poiss01}\\
  \mathcal{P}_2&(\om(z),\hat{\om}(z))\nn\\
  =&\Big(-a(z)(\{a(z),\om(z)\}+\{\hat{a}(z),\hat{\om}(z)\})_-
  +\{a(z),(a(z)\om(z)+\hat{a}(z)\hat{\om}(z))_-\},
  \nn\\
  & \quad \hat{a}(z)(\{a(z),\om(z)\}+\{\hat{a}(z),\hat{\om}(z)\})_+
  -\{\hat{a}(z),(a(z)\om(z)+\hat{a}(z)\hat{\om}(z))_+\}\Big).\label{poiss02}
  \end{align}
They give Poisson brackets $\{\ ,\ \}_{1,2}$ that represent the
dispersionless two-component BKP hierarchy \eqref{disphir01},
\eqref{disphir02} to a bi-Hamiltonian form as
\begin{align}\label{Ham2BKP}
\frac{\p F}{\p s_k}=\{F, H_{k+2m}\}_1=\{F, H_{k}\}_2, \quad \frac{\p
F}{\p \hat{s}_k}=\{F, \hat{H}_{k+2n}\}_1=\{F, \hat{H}_{k}\}_2
\end{align}
with  $k\in\mathZ_+^{odd}$ and
\begin{align}
  H_k=\frac{2m}{k}\frac{1}{2\pi\bmi}\oint_{S^1}\oint_{|z|=1}\lm(z)^{k}d z\,d x,\
  \hat{H}_k=\frac{2n}{k}\frac{1}{2\pi\bmi}\oint_{S^1}\oint_{|z|=1}\hat{\lm}(z)^{k}d z\,d x.
\end{align}
\end{prop}

Introduce two generating functions for local functionals as
\begin{equation}
 a(p,y)=p^{2m}+\sum_{i\leq m}v_i(y) p^{2i-2}, \quad
   \hat{a}(p,y)=\sum_{j\geq -n}\hat{v}_j(y) p^{2j}.
\end{equation}
Their gradients are (cf. \eqref{rpgener01}, \eqref{rpgener02})
  \begin{align}\label{oneform}
  & da(p,y)=\Big(\frac{p^{2m}}{z^{2m-1}(p^2-z^2)}\dt(x-y),0\Big),\quad
  |z|<|p|;
  \\
   & d\hat{a}(p,y)=\Big(0,\frac{z^{2n+1}}{p^{2n}(z^2-p^2)}\dt(x-y)\Big),\quad
  |z|>|p|.
  \end{align}
With the help of these notations, the Poisson brackets
 \eqref{poiss01} and \eqref{poiss02} can be expressed in the form
 \begin{align}
 \label{bh1}
  \{\al(p,x),\beta(q,y)\}_1=&\Big[\frac{q\p_q\beta(q,x)}{q^2-p^2}
  +\frac{p\p_p\al(p,x)}{p^2-q^2}\Big]\dt'(x-y)\nn\\
  &+\Big[\frac{p^2+q^2}{(p^2-q^2)^2}\big(\p_x\al(p,x)-\p_x\beta(q,x)\big)\Big]\dt(x-y),
\\
  \{\al(p,x),\beta(q,y)\}_2=&\Big[\frac{q\p_q\beta(q,x)}{q^2-p^2}\al(p,x)
  +\frac{p\p_p\al(p,x)}{p^2-q^2}\beta(q,x)\Big]\dt'(x-y)\nn\\
  &+\Big[\frac{p^2+q^2}{(p^2-q^2)^2}\big(\p_x\al(p,x)\cdot\beta(q,x)
  -\al(p,x)\p_x\beta(q,x)\big)\nn\\
  &+\frac{p\p_p\al(p,x)\cdot\p_x\beta(q,x)-q\al(p,x)\p_q\p_x\beta(q,x)}{p^2-q^2}\Big]\dt(x-y),
\label{bh2}
  \end{align}
  where $\al,\beta\in\{a,\hat{a}\}$.
These Poisson brackets are of hydrodynamic type.

Comparing the coefficients of $\dt'(x-y)$ in \eqref{bh1},
\eqref{bh2} with the metrics \eqref{cometric} and \eqref{iinters2}
on the Frobenius manifold $\cM_{m,n}$, we have the following result.
\begin{prop}
  The bi-Hamiltonian structure associated to the Frobenius manifold
$\cM_{m,n}$ is given by \eqref{bh1} and \eqref{bh2} for the
dispersionless two-component BKP hierarchy.
\end{prop}

To obtain the principal hierarchy associated to $\cM_{m,n}$, let us
introduce the Hamiltonians
\begin{equation}\label{ham}
\mathH_{u,p-1}=\int \ta_{u,p}\,d x, \quad p\ge 0
\end{equation}
with
\begin{equation}\label{theta}
 \ta_{u,p}=\left\{ \begin{aligned}
    &\frac{1}{2 i+1}\frac{1}{(2p) !!}\frac{1}{2\pi \mathbbm{i}}
    \oint_{|z|=1}\zeta(z)^{\frac{2i+1}{2}}\varphi(z)^p\,d z,
      &&u=t^i~ (i\in\mathbb{Z}); \\
       \\
    &-\frac{\Gamma\left(\frac{2m-2j+1}{2m}\right)}{2m\,\Gamma\left(p+1+\frac{2m-2j+1}{2m}\right)}
    \res_{z=\infty} a(z)^{p+\frac{2m-2j+1}{2m}}\,d z, &&u=h^j ~(1\le j\le m); \\
    \\
    &\frac{\Gamma\left(\frac{2n-2k+1}{2n}\right)}{2n\,\Gamma\left(p+1+\frac{2n-2k+1}{2n}\right)}
    \res_{z=0} \hat{a}(z)^{p+\frac{2n-2k+1}{2n}}\,d z, &&u=\hat{h}^k ~(1\le k\le n).
          \end{aligned} \right.
 \end{equation}
Here $\zeta(z)=a(z)-\hat{a}(z)$ and $\varphi(z)=a(z)+\hat{a}(z)$.

\begin{thm}\label{thm-princ}
  The principal hierarchy associated to the
  Frobenius manifold $\cM_{m,n}$ consists of the following bi-Hamiltonian
  flows
  \begin{align}\label{biham}
  \frac{\p F}{\p T^{u,p}}=&\big\{F,\mathH_{u,p}\big\}_1
  =\left(p+\frac{1}{2}+\mu_{u}\right)^{-1}\big\{F,\mathH_{u,p-1}\big\}_2,\quad p\geq
  0,
  \end{align}
  where
  \begin{equation}
 \mu_{u}=\left
 \{ \begin{aligned}
    &i,   &&u=t^i~ (i\in\mathbb{Z}); \\
    &\frac{m-2j+1}{2m}, &&u=h^j~(1\le j\le m);\\
    &\frac{n-2k+1}{2n}, &&u=\hat{h}^k~(1\le k\le n).
          \end{aligned} \right.
 \end{equation}
\end{thm}
\begin{prf}
In comparison with \eqref{flatc01}--\eqref{flatc03}, we have
\[
\ta_{t^i,0}=-\frac{1}{2} t^{-i}, \quad  \ta_{h^j,0}=\frac{1}{2m}
h^{m+1-j}, \quad \ta_{\hat{h}^k,0}= \frac{1}{2n} \hat{h}^{n+1-k},
\]
which are densities of Casimirs for the first Poisson bracket (cf.
\eqref{recursion}). According to the theory of Frobenius manifold
and principal hierarchy \cite{Du, DZ}, we only need to check the
recursion relation \eqref{biham}.

Take $u=t^i$ for example. It is easy to compute the gradients at
point $(a(z),\hat{a}(z))$:
\begin{align*}\label{}
d\mathH_{t^i,p-1}=&\frac{1}{(2i+1)\,(2p)!!}
\left(\left(\frac{2i+1}{2}\zeta(z)^{\frac{2i-1}{2}}\varphi(z)^p+ p\,
\zeta(z)^{\frac{2i+1}{2}}\varphi(z)^{p-1}\right)_{\ge -2 m+1},
\right.
\\
&\qquad\qquad
\left.\left(-\frac{2i+1}{2}\zeta(z)^{\frac{2i-1}{2}}\varphi(z)^p+
p\, \zeta(z)^{\frac{2i+1}{2}}\varphi(z)^{p-1}\right)_{\le 2 n-1}
\right).
\end{align*}
Substituting them into \eqref{poiss01} and \eqref{poiss02}, we have
\begin{align}\label{}
& \mathcal{P}_1(d\mathH_{t^i,p})=\left(\left\{a(z),
(A_{t^i,p}(z))_-\right\},
  \left\{ (A_{t^i,p}(z))_+, \hat{a}(z)\right\}\right),
  \\
&
\mathcal{P}_2(d\mathH_{t^i,p-1})=\left(p+i+\frac{1}{2}\right)\left(\left\{a(z),
(A_{t^i,p}(z))_-\right\},
  \left\{ (A_{t^i,p}(z))_+, \hat{a}(z)\right\}\right)
\end{align}
with
\[
A_{t^i,p}(z)=\frac{1}{2
i+1}\frac{1}{(2p)!!}\zeta(z)^{\frac{2i+1}{2}}\varphi(z)^p.
\]
The other cases are similar. The theorem is proved.
\end{prf}

\begin{prfn}{Theorem~\ref{prinhier}} This is a corollary of
Theorem~\ref{thm-princ}.
\end{prfn}

\begin{rem}
To obtain the primary flows $\p/\p T^{u,0}$ of the principal
hierarchy, one can also use the formula \eqref{prflow} and the
$3$-point functions computed in Section~2.3. The result coincides
with that in theorem~\ref{prinhier}.
\end{rem}

Observe that the principal hierarchies with different $m$ and $n$
have their common part as the  dispersionless two-component BKP
hierarchy. In fact, comparing \eqref{principal} with
\eqref{disphir01} and \eqref{disphir02} one has
\begin{align*}\label{}
& T^{h^j,p}=
\frac{2m\,\Gamma\left(p+1+\frac{2m-2j+1}{2m}\right)}{\Gamma\left(\frac{2m-2j+1}{2m}\right)}s_{2m
p+2m-2j+1}, \quad 1\le j \le m; \\
& T^{\hat{h}^k,p}=
\frac{2n\,\Gamma\left(p+1+\frac{2n-2k+1}{2n}\right)}{\Gamma\left(\frac{2n-2k+1}{2n}\right)}
\hat{s}_{2n p+2n-2k+1}, \quad 1\le k \le n.
\end{align*}
\begin{defn}
The hierarchy \eqref{principal} (or equivalently \eqref{biham}) is
called the $(2m,2n)$-principal two-component BKP hierarchy.
\end{defn}

We end this section with two remarks on the principal two-component
BKP hierarchy.
\begin{rem}
Recall the canonical coordinates $u_{\sg}$ in Section~2.4. The
principal hierarchy \eqref{principal} can also be represented in the
following linear form:
  \begin{align}
  \frac{\p u_{\sg}}{\p T^{u,p}}=\left.\mathcal{A}_{u,p}(z)
  \right|_{z=z(\sg):\,\Sigma\to S^1}\frac{\p u_{\sg}}{\p x},
  \end{align}
  where
\begin{equation*}
 \mathcal{A}_{u,p}(z)=\left
 \{ \begin{aligned}
    &\left(A_{u,p}(z)\Big(\frac{2i+1}{2\zeta(z)}+\frac{p}{\varphi(z)}\Big)\frac{\p a(z)}{\p z}\right)_+
     &&\\
&     \quad
+\left(A_{u,p}(z)\Big(\frac{2i+1}{2\zeta(z)}-\frac{p}{\varphi(z)}\Big)\frac{\p
\hat{a}(z)}{\p
    z}\right)_-, \quad
    &&u=t^i\ (i\in\mathZ);\\
    &\left(\frac{\p A_{u,p}(z)}{\p z}\right)_+,
     && u=h^j\ (1\le j\le m); \\
    &-\left(\frac{\p A_{u,p}(z)}{\p z}\right)_-,
     && u=\hat{h}^k\ (1\le k\le n)
          \end{aligned}
          \right.
 \end{equation*}
with $A_{u,p}$ given in \eqref{Aup}.
\end{rem}

\begin{rem}
In \cite{CT} Chen and Tu rewrote the dispersionless two-component
BKP hierarchy to a system of dispersionless Hirota equations of a
free energy $\cF$. Given any solution $\cF$ of such Hirota
equations, one introduces
\[
\cF_{i,j,k}=\p_{s_i}\p_{s_j}\p_{s_k}\cF, \quad i, j,
k\in\mathZ^{odd}
\]
with $s_{-k}=\hat{s}_{k}$ for $k\in\mathZ^{odd}_+$, and defines
structure constants $C_{i,j}^l$ such that
\[
\cF_{i,j,k}=\sum_{l\in\mathZ^{odd}}C_{i,j}^l \cF_{k,l,1}, \quad i,
j, k\in\mathZ^{odd}.
\]
Then $\cF$ satisfies the following WDVV equations \cite{CT}
\begin{equation} \label{WDVV-CT}
\sum_{l\in\mathZ^{odd}}C_{i,j}^l\cF_{l,k,h}=\sum_{l\in\mathZ^{odd}}C_{i,k}^l\cF_{l,j,h},\quad
i, j, k, h\in\mathZ^{odd}.
\end{equation}

Observe that the associativity equations \eqref{WDVV-CT} look
similar with those satisfied by the potential of the Frobenius
manifolds $\cM_{m,n}$. In order to clarify the relation between
them, it probably needs to develop a theory of topological solution
for the principal hierarchy associated to infinite-dimensional
Frobenius manifold, as was discussed in the last section of
\cite{CDM}

\end{rem}

\section{Finite-dimensional Frobenius submanifolds}

The manifold $\mathM_{m,n}$ can be represented as $\cM_{m,n}^0\times
M_{m,n}$. Here $\cM_{m,n}^0$ and $M_{m,n}$ are submanifolds spanned
by the flat coordinates $\{t^i\mid i\in\mathZ\}$ and
$\{h^j\}_{j=1}^m\cup\{\hat{h}^k\}_{k=1}^n$ respectively. In other
words, they are characterized by the functions $\zeta(z)$ and $l(z)$
given in \eqref{funcpair} respectively.

Let us consider the $(m+n)$-dimensional submanifold $M_{m,n}$. For
convenience its coordinates are redenoted as (recall \eqref{flatc02}
and \eqref{flatc03})
\begin{equation}\label{flatw}
w^\al=\left\{\begin{aligned}
    & h^{m+1-\al},
     && 1\le\al\le m; \\
     & \hat{h}^{m+n+1-\al}, && m+1\le\al\le m+n.
     \end{aligned}
\right.
\end{equation}
The projection $\mathM_{m,n} \to M_{m,n}$ induces a Frobenius
structure on $M_{m,n}$. After some straightforward calculation the
following result is obtained (cf. Lemma~4.5 in \cite{Du}).
\begin{prop}
The submanifold $M_{m,n}$ is a semisimple Frobenius manifold with
potential $F_{m,n}$ given in \eqref{poten03}. Let $\p_1$, $\p_2$ and
$\p_3$ denote arbitrary tangent vector fields on this manifold.
\begin{itemize}
  \item[(i)] The invariant inner product (flat metric) $<\ ,\ >$ and
  the $3$-tensor $c$ are
  \begin{align}\label{}
  <\p_1,\p_2>=&\res_{l'(z)=0}\frac{\p_1l(z)\cdot\p_2l(z)}{l'(z)}dz, \\
  c(\p_1,\p_2,\p_3)=&\res_{l'(z)=0}\frac{\p_1 l(z)\cdot\p_2 l(z)\cdot\p_3
l(z)}{l'(z)}d z.
  \end{align}
A system of flat coordinates for the metric $<\ ,\ >$ is given by
$\{w^\al\}$.
  \item[(ii)]  The unity vector field is $e=\p/\p w^1$.
  \item[(iii)]
The Euler vector field is
\begin{equation}\label{}
E_{m,n}=\sum_{\al=1}^m\frac{m-\al+1}{m}w^\al\frac{\p}{\p w^\al}
   +\sum_{\al=m+1}^{m+n}\left(\frac{2(m+n-\al)+1}{2n}+\frac{1}{2m}\right)w^\al\frac{\p}{\p w^\al}.
\end{equation}
\item[(iv)] Suppose $l'(z)=0$ has $2(m+n)$ pairwise distinct
roots $\pm z_1, \pm z_2, \dots, \pm z_{m+n}\in\mathbb{C}$. Then the
canonical coordinates defined by
\begin{equation}\label{}
u_i=l(z_i)=l(-z_i), \quad i=1, 2, \dots, m+n
\end{equation}
satisfy
\begin{equation}\label{}
\frac{\p}{\p u_i}\cdot\frac{\p}{\p u_j}=\dt_{ij}\frac{\p}{\p u_i},
\quad <\frac{\p}{\p u_i}, \frac{\p}{\p
u_j}>=\dt_{ij}\frac{2}{l''(z_i)}.
\end{equation}
\item[(iv)] The intersection form reads
\begin{equation}\label{}
(\p_1,\p_2)=\res_{l'(z)=0}\frac{\p_1 \log l(z)\cdot\p_2 \log
l(z)}{\p_z \log l(z)}dz
\end{equation}
\end{itemize}
\end{prop}

Observe that the Frobenius manifold $M_{m,n}$ coincides with the one
defined on the orbit space of the Coxeter group $B_{m+n}$ by Bertola
(see Appendix~B in \cite{Bertola}) starting from a superpotential of
the form $l(z)$. In particular, the manifold $\mathcal{M}_{m,1}$ is
defined on the orbit space $\mathbb{C}^{m+1}/D_{m+1}$ of the Coxeter
group $D_{m+1}$. The Frobenius manifolds $M_{m,n}$ were
reconstructed by Zuo \cite{Zuo}, whose method is based on the
polynomial contravariant metric of the orbit space induced from the
Euclidean metric (see for example Lemma~4.1 in \cite{Du}) and
different choices of unity vector field.

As in \cite{Bertola, Zuo}, one sees that in general the potential
$F_{m,n}$ for $M_{m,n}$ is a polynomial in $w^1,w^2,\dots,w^{m+n},
1/w^{m+n}$, and is a polynomial in $w^1,w^2,\dots,w^{m+n}$ if and
only if $n=1$. Following are the potentials for small $m$ and $n$.
\begin{exam} For $n=1$,
\begin{align}
F_{1,1}=&w^1\left(\frac{(w^1)^2}{12}+\frac{(w^2)^2}{4}\right), \\
F_{2,1}=&w^1\left(\frac{w^1
w^2}{8}+\frac{(w^3)^2}{4}\right)+\frac{(w^2)^5}{3840}+\frac{(w^2)^2(w^3)^2}{32}, \\
F_{3,1}=&w^1\left(\frac{w^1
w^3+(w^2)^2}{12}+\frac{(w^4)^2}{4}\right)+\frac{(w^2)^2(w^3)^3}{1296}-\frac{(w^2)^3
w^3}{216}+\frac{(w^3)^7}{1632960}\nn\\
&+\left(\frac{w^2 w^3}{24}+\frac{(w^3)^3}{432}\right)(w^4)^2.
\end{align}
For $n>1$,
\begin{align}
F_{1,2}=&w^1\left(\frac{(w^1)^2}{12}+\frac{w^2
w^3}{4}\right)+\frac{(w^3)^4}{768}+\frac{(w^2)^3}{6\,w^3},
\\
F_{2,2}=&w^1\left(\frac{w^1 w^2}{8}+\frac{w^3
w^4}{4}\right)+\frac{(w^2)^5}{3840}+\frac{(w^2)^2 w^3
w^4}{32}+\frac{w^2 (w^4)^4}{768}+\frac{(w^3)^3}{6\,w^4},
\\
F_{1,3}=&w^1\left(\frac{(w^1)^2}{12}+\frac{w^2
w^4}{6}+\frac{(w^3)^2}{12}\right)+\frac{w^3
(w^4)^3}{648}+\frac{(w^2)^2 w^3}{2\,w^4}-\frac{w^2
(w^3)^3}{3\,(w^4)^2}+\frac{(w^3)^5}{10\,(w^4)^3}.
\end{align}
\end{exam}
In particular, the potential $F_{m,1}$ coincides with the partition
function that solves the $D_{m+1}$-model in topological field theory
\cite{DVV} up to a rescaling of the flat coordinates.

The projection from $\mathM_{m,n}$ to its submanifold $M_{m,n}$ can
be realized by setting $a(z)=\hat{a}(z)$ so that
 $\zeta(z)=a(z)-\hat{a}(z)$ vanishes. Such a constraint
 corresponds to the $(2\,m,2\,n)$-reduction of the associated
 bi-Hamiltonian structure \eqref{poiss01}, \eqref{poiss02}, see
 \cite{Wu} for details. Hence by using Theorem~\ref{prinhier} we
 have the following result.
\begin{prop}
The principal hierarchy associated to $M_{m,n}$ is the
$(2\,m,2\,n)$-reduction of the dispersionless two-component BKP
hierarchy, i.e.,
\begin{equation}\label{prl}
\frac{\p l(z)}{\p T^{\al,p}}=\{A_{\al,p}(z),\, l(z)\}, \quad
1\le\al\le m+n \hbox{ and } p\ge0,
\end{equation}
where
\begin{equation*}
A_{\al,p}(z)=\left\{\begin{array}{ll}
      \dfrac{\Gm\left(\frac{2\al-1}{2m}\right)}
{2\,m\,\Gm\left(p+1+\frac{2\al-1}{2m}\right)}\left(l(z)^{p+\frac{2\al-1}{2m}}\right)_+,
\quad & 1\le\al\le
m; \\ \\
-\dfrac{\Gm\left(\frac{2(\al-m)-1}{2n}\right)}
{2\,n\,\Gm\left(p+1+\frac{2(\al-m)-1}{2n}\right)}\left(l(z)^{p+\frac{2(\al-m)-1}{2n}}\right)_-,
\quad & m+1\le\al\le m+n
     \end{array}
\right.
\end{equation*}
with $l(z)^{1/2m}=z+O(z^{-1})$ as $z\to\infty$ and
$l(z)^{1/2n}=w^{m+n}\,z^{-1}/2n+O(z)$ as $z\to0$.
\end{prop}

Particularly when $n=1$, the hierarchy \eqref{prl} is the
dispersionless limit of the Drinfeld-Sokolov hierarchy associated to
Kac-Moody algebra $D^{(1)}_{m+1}$ and the vertex $c_0$ of its Dynkin
diagram, see \cite{DS, LWZ} for details.

Therefore, Proposition~\ref{prp-mn} is proved.

If we constrain $l(z)$ by $l(z)_-=0$, then the flows $\p/\p
T^{\al,p}$ in \eqref{prl} with $1\le \al\le m$ are well defined, and
they compose the dispersionless limit of the Drinfeld-Sokolov
hierarchy of type $(B_m^{(1)},c_0)$. This observation coincides with
the fact that the polynomial potential for the Frobenius manifold on
the orbit space of Coxeter group $B_m$ is obtained by omitting the
terms containing $w^{\al}$ with $\al>m$ in $F_{m,n}$.

\begin{rem}
It is easy to see that the infinite-dimensional submanifold
$\cM_{m,n}^0$ of $\cM_{m,n}$ fulfills all conditions to define a
Frobenius manifold except the existence of a unity vector field. A
manifold of this kind is called a \emph{quasi-Frobenius manifold}
following \cite{DZ} (see Section~3.2 there). Such manifolds will be
studied in a follow-up work.
\end{rem}

\section{Conclusion and outlook}

For every pair of positive integers $m$ and $n$, we have constructed
an infinite-dimensional semisimple Frobenius manifold $\cM_{m,n}$ on
the space that consists of pairs of meromorphic even functions
$\big(a(z),\hat{a}(z)\big)$ satisfying the conditions (M1)-(M3).
Such Frobenius manifolds have rich geometric properties described in
a similar way as in the finite-dimensional case.

The bi-Hamiltonian structure \eqref{bh1}, \eqref{bh2} of
hydrodynamic type associated to the Frobenius manifold $\cM_{m,n}$
is just the one for the dispersionless two-component BKP hierarchy.
Based on this fact, we have obtained the Lax representation
\eqref{principal} of the principal hierarchy for $\cM_{m,n}$, which
is a wide extension of the dispersionless two-component BKP
hierarchy. Note that the definition of the complementary flows
$\p/\p T^{t^i,p}$ relies on the analyticity property of the
functions $a(z)$ and $\hat{a}(z)$. How to deform these flows, such
as from $\p/\p T^{h^j,p}$ and $\p/\p T^{\hat{h}^k,p}$ to the full
flows of the two-component BKP hierarchy, has not been considered.
From another point of view, a tau function of the principal
hierarchy can be defined by using the tau-symmetric Hamiltonian
densities $\ta_{\al,p}$ in \eqref{theta}. Whether the tau function
has a topological deformation, which might be governed by an
analogue of the universal loop equation in \cite{DZ}, will be
studied in subsequent publications.

We have shown that $\cM_{m,n}$ contains an $(m+n)$-dimensional
Frobenius submanifold $M_{m,n}$, which coincides with the semisimple
Frobenius manifold defined on the orbit space of the Coxeter group
$B_{m+n}$ (or $D_{m+1}$ when $n=1$). Moreover, the principal
hierarchy for $M_{m,n}$ is shown to be the $(2\, m,2\, n)$-reduction
of the dispersionless two-component BKP hierarchy. In particular,
when $n=1$, the principal hierarchy is just the dispersionless limit
of Drinfeld-Sokolov hierarchy of type $(D_{m+1}^{(1)},c_0)$. It is
interesting to look for infinite-dimensional Frobenius manifolds
that contain Frobenius submanifolds defined on the orbit space of
Coxeter groups besides types B and D, and relate them to integrable
systems such as Drinfeld-Sokolov hierarchies.

Finally, we hope that the present result would be helpful to
achieving a normal form for infinite-dimensional Frobenius manifolds
underlying $2+1$ integrable systems. For instance, the technique
here probably helps to generalize the result in \cite{CDM} such that
there is also a class of Frobenius manifolds for the dispersionless
2D Toda hierarchy, and they possess submanifolds of finite dimension
associated with the extended bigraded Toda hierarchy \cite{CDZ,
Carlet}.

\vskip 0.5truecm \noindent{\bf Acknowledgments.} The authors are
grateful to Guido Carlet, Boris Dubrovin, Si-Qi Liu and Youjin Zhang
for helpful discussions. The first named author acknowledges the
support of the Young SISSA Scientist Grant ``FIMA.645''. The second
named author is partially supported by the National Basic Research
Program of China (973 Program)  No.2007CB814800, the NSFC
No.10801084 and 11071135; he is also supported by the State
Scholarship Fund of China No.2010621203 and would like to express
thanks for the hospitality of SISSA where part of the work was done.


\begin{appendices}
\renewcommand{\thesection}{A\Alph{section}}

\section*{Appendix: Proof of Lemma~\ref{lem-finF}}

\begin{prfn}{Lemma~\ref{lem-finF}}
For any $u,v,w,s\in \bm{h}\cup\hat{\bm{h}}$, let
\begin{equation}\label{}
C_{uvw}=-\big(\res_{\infty}+\res_{0}\big)\frac{\p_{u}
l(z)\cdot\p_{v}
    l(z)\cdot\p_{w} l(z)}{l'(z)}d z,
\end{equation}
then
\begin{align}\label{Cuvws}
\p_s C_{uvw}=&-\big(\res_{\infty}+\res_{0}\big)\frac{ \p_s\p_{u}
l(z)\cdot\p_{v}
    l(z)\cdot\p_{w} l(z)+\mathrm{c.p.}(u,v,w) }{l'(z)}d z
    \nn\\
&+\big(\res_{\infty}+\res_{0}\big)\frac{\p_{u} l(z)\cdot\p_{v}
    l(z)\cdot\p_{w} l(z)\cdot \p_z \p_s l(z)}{l'(z)^2}  d z.
\end{align}
Here $\mathrm{c.p.}$ denotes the cyclic permutation. To show the
lemma, it only needs to check that \eqref{Cuvws} is symmetric with
respect to $u, v, w$ and $s$. We will employ the formulae
\eqref{formu02}, \eqref{formu03} and \eqref{dlh} to do this.

First, let us compute $(2m)^2\p_{h^{j_4}} C_{h^{j_1} h^{j_2}
h^{j_3}}$. By using \eqref{formu02} and \eqref{dlh} we have
\begin{equation}\label{}
\frac{\p l(z)}{\p h^j}=\Big(\chi(z)^{2m-2j}\chi'(z)\Big)_+, \quad
\frac{\p^2 l(z)}{\p h^{j_1} \p
h^{j_2}}=\frac1{2m}\p_z\Big(\chi(z)^{2m-2j_1-2j_2+1}\chi'(z)\Big)_+.
\end{equation}
In this case the first term on the right hand side of \eqref{Cuvws}
yields
\begin{align}
& -(2m)^2 \big(\res_{\infty}+\res_{0}\big)\frac{ \p_{h^{j_4}}
\p_{h^{j_1}} l(z)\cdot\p_{h^{j_2}}
    l(z)\cdot\p_{h^{j_3}} l(z)}{l'(z)}d z
    \nn\\
=&-\res_{\infty}\frac{ \p_z \big(\chi^{2m-2j_1-2j_4+1}\chi'\big)_+
\cdot \big(\chi^{2m-2j_2}\chi'\big)_+\big(\chi^{2m-2j_3}\chi'\big)_+
}{\chi^{2m-1}\chi'}d z
\nn\\
=&-\res_{\infty}\frac{ \p_z
\big(\chi^{2m-2j_1-2j_4+1}\chi'\big)_+}{\chi^{2m-1}\chi'} \Big(
\chi^{4m-2j_2-2j_3}(\chi')^2 \nn\\
&\qquad
-\big(\chi^{2m-2j_2}\chi'\big)_\oplus\big(\chi^{2m-2j_3}\chi'\big)_-
-\big(\chi^{2m-2j_2}\chi'\big)_-\big(\chi^{2m-2j_3}\chi'\big)_\oplus
\Big)_\oplus d z
\nn\\
=&-\res_{\infty}\p_z \big(\chi^{2m-2j_1-2j_4+1}\chi'\big)_+ \cdot
\chi^{2m-2j_2-2j_3+1}\chi' d z \nn\\
&-\res_{\infty}\p_z \big(\chi^{2m-2j_1-2j_4+1}\chi'\big)_\oplus
\cdot\Big(-\chi^{-2j_2+1}\big(\chi^{2m-2j_3}\chi'\big)_-
-\big(\chi^{2m-2j_2}\chi'\big)_-\chi^{-2j_3+1}\Big) d z
\nn\\
=&-\res_{\infty}\p_z \big(\chi^{2m-2j_1-2j_4+1}\chi'\big)_+
\chi^{2m-2j_2-2j_3+1}\chi'd z \nn\\
&-\res_{\infty}\Big(\chi^{2m-2j_1-2j_2-2j_4+2}\chi'\p_z\big(\chi^{2m-2j_3}\chi'\big)_-
+\chi^{2m-2j_1-2j_3-2j_4+2}\chi'\p_z\big(\chi^{2m-2j_2}\chi'\big)_-
\Big) d z
\nn\\
&-\res_{\infty}\Big((-2j_2+1)\chi^{2m-2j_1-2j_2-2j_4+2}(\chi')^2\big(\chi^{2m-2j_3}\chi'\big)_-
\nn\\
&\qquad
+(-2j_3+1)\chi^{2m-2j_1-2j_3-2j_4+2}(\chi')^2\big(\chi^{2m-2j_2}\chi'\big)_-
\Big) d z. \label{term1}
\end{align}
Here and below the subscript ``$\oplus$'' means that the operation
of taking the nonnegative part of the series can be omitted. On the
right hand side of \eqref{term1} the first term can be rewritten to
\begin{align}\label{}
&\res_{\infty}\chi^{2m-2j_1-2j_4+1}\chi' \p_z
\big(\chi^{2m-2j_2-2j_3+1}\chi'\big)_-d z
 \nn\\
=& \res_{\infty}\chi^{2m-2j_1-2j_4+1}\chi' \p_z
\big(\chi^{2m-2j_2-2j_3+1}\chi'-\big(\chi^{2m-2j_2-2j_3+1}\chi'\big)_+
\big)d z, \nn
\end{align}
hence
\begin{align}\label{}
&-\res_{\infty}\p_z \big(\chi^{2m-2j_1-2j_4+1}\chi'\big)_+
\chi^{2m-2j_2-2j_3+1}\chi' \nn\\
=& \frac1{2}\res_{\infty}\Big((2m-2j_2-2j_3+1)
\chi^{2m-2j_1-2j_2-2j_3-2j_4+1}(\chi')^3 \nn\\
& \qquad +\chi^{2m-2j_1-2j_2-2j_3-2j_4+2}\chi'\chi''\Big)d z
 \nn\\
&-\frac1{2}\res_{\infty}\Big(\p_z
\big(\chi^{2m-2j_1-2j_4+1}\chi'\big)_+\cdot
\chi^{2m-2j_2-2j_3+1}\chi' \nn\\
&\qquad + \chi^{2m-2j_1-2j_4+1}\chi' \p_z
\big(\chi^{2m-2j_2-2j_3+1}\chi'\big)_+  \Big)d z. \label{term11}
\end{align}
On the other hand, $(2m)^2$ times of the last term of \eqref{Cuvws}
equals
\begin{align}
&\res_{\infty}\frac{ \big(\chi^{2m-2j_1}\chi'\big)_+
\big(\chi^{2m-2j_2}\chi'\big)_+\big(\chi^{2m-2j_3}\chi'\big)_+\p_z
\big(\chi^{2m-2j_4}\chi'\big)_+ }{(\chi^{2m-1}\chi')^2}d z \nn\\
=&\res_{\infty}\frac{ \big(\chi^{2m-2j_3}\chi'\big)_+\p_z
\big(\chi^{2m-2j_4}\chi'\big)_+}{(\chi^{2m-1}\chi')^2} \Big(
\chi^{4m-2j_1-2j_2}(\chi')^2 \nn\\
&\qquad
-\big(\chi^{2m-2j_1}\chi'\big)_\oplus\big(\chi^{2m-2j_2}\chi'\big)_-
-\big(\chi^{2m-2j_1}\chi'\big)_-\big(\chi^{2m-2j_2}\chi'\big)_\oplus
\Big)_\oplus d z
\nn\\
=&\res_{\infty}\chi^{-2j_1-2j_2+2}\big(\chi^{2m-2j_3}\chi'\big)_+\p_z
\big(\chi^{2m-2j_4}\chi'\big)_+\, d z \nn\\
&-\res_{\infty}\frac{\big(\chi^{2m-2j_3}\chi'\big)_\oplus\p_z
\big(\chi^{2m-2j_4}\chi'\big)_+}{\chi^{2m-1}\chi'}
\Big(\chi^{-2j_1+1}\big(\chi^{2m-2j_2}\chi'\big)_-
\nn\\
&\qquad +\big(\chi^{2m-2j_1}\chi'\big)_-\chi^{-2j_2+1}\Big) d z
\nn\\
=&\res_{\infty}\chi^{-2j_1-2j_2+2}\Big(\chi^{2m-2j_3}\chi'\p_z
\big(\chi^{2m-2j_4}\chi'\big) \nn\\
&\qquad - \big(\chi^{2m-2j_3}\chi'\big)_\oplus\p_z
\big(\chi^{2m-2j_4}\chi'\big)_- -\big(\chi^{2m-2j_3}\chi'\big)_-\p_z
\big(\chi^{2m-2j_4}\chi'\big)_\oplus\Big)_\oplus \, d z \nn\\
&+
\res_{\infty}\p_z\Big(\chi^{-2j_1-2j_3+2}\big(\chi^{2m-2j_2}\chi'\big)_-
+\chi^{-2j_2-2j_3+2}\big(\chi^{2m-2j_1}\chi'\big)_-\Big)\cdot
\chi^{2m-2j_4}\chi' d
z \nn\\
=&\res_{\infty}\Big((2m-2j_4)\chi^{4m-2j_1-2j_2-2j_3-2j_4+1}(\chi')^3
+\chi^{4m-2j_1-2j_2-2j_3-2j_4+2}\chi' \chi''\Big)d z \nn\\
&-\res_{\infty}\chi^{2m-2j_1-2j_2-2j_3+2}\chi'\p_z
\big(\chi^{2m-2j_4}\chi'\big)_-d z
\nn\\
&+\res_{\infty}\Big(\chi^{2m-2j_2-2j_3-2j_4+2}\chi'\p_z
\big(\chi^{2m-2j_1}\chi'\big)_-+\mathrm{c.p.}(j_1,j_2,j_3) \Big)d z \nn\\
&+\res_{\infty}\Big((-2j_2-2j_3+2)\chi^{2m-2j_2-2j_3-2j_4+1}(\chi')^2
\big(\chi^{2m-2j_1}\chi'\big)_-+\mathrm{c.p.}(j_1,j_2,j_3) \Big)d z.
\label{term2}
\end{align}
Substituting \eqref{term1}--\eqref{term2} into \eqref{Cuvws} we
obtain
\begin{align}\label{}
&(2m)^2\p_{h^{j_4}} C_{h^{j_1} h^{j_2} h^{j_3}} \nn\\
&=\res_{\infty}\bigg(\Big(5m-2j_1-2j_2-2j_3-2j_4+\frac{3}{2}\Big)
\chi^{4m-2j_1-2j_2-2j_3-2j_4+1}(\chi')^3 \nn\\
&\qquad +\frac{5}{2}\chi^{4m-2j_1-2j_2-2j_3-2j_4+2}\chi'
\chi''\bigg)d z
\nn\\
&-\res_{\infty}\Big(\chi^{2m-2j_2-2j_3-2j_4+2}\chi'\p_z
\big(\chi^{2m-2j_1}\chi'\big)_-+\mathrm{c.p.}(j_1,j_2,j_3,j_4)
\Big)d z
\nn\\
&-\frac1{2}\res_{\infty}\Big(\p_z
\big(\chi^{2m-2j_1-2j_4+1}\chi'\big)_+\cdot
\chi^{2m-2j_2-2j_3+1}\chi' \nn\\
&\qquad + \chi^{2m-2j_1-2j_4+1}\chi' \p_z
\big(\chi^{2m-2j_2-2j_3+1}\chi'\big)_+ +\mathrm{c.p.}(j_1,j_2,j_3)
\Big)d z,
\end{align}
which is indeed symmetric with respect to
$j_1,j_2,j_3,j_4\in\{1,2,\dots,m\}$.

For $u,v,w,s\in\hat{\bm{h}}$ it is almost the same. With similar
method, the other cases are easy to check by virtue of the fact
$\p_{h^j}\p_{\hat{h}^k}l(z)=0$. Therefore Lemma~\ref{lem-finF} is
proved.

\end{prfn}

\end{appendices}

\end{document}